\documentclass{article}
\usepackage{epsfig}
\usepackage{latexsym}
\begin{document}

\begin{titlepage}
\vspace{0.75cm}
\title{Hadronic Parity Violation: an Analytic Approach}
\author{Barry R. Holstein\\
Department of Physics-LGRT\\
University of Massachusetts\\
Amherst, MA  01003} \maketitle

\begin{abstract}

Using a recent reformulation of  the analysis of nuclear
parity-violation (PV) using the framework of effective field
theory (EFT), we show how predictions for parity-violating
observables in low energy light hadronic systems can be understood
in an analytic fashion.  It is hoped that such an analytic
approach may encourage additional experimental work as well as add
to the understanding of such parity-violating phenomena, which is
all too often obscured by its description in terms of numerical
results obtained from complex two-body potential codes.
\end{abstract}
\vfill
\end{titlepage}

%%%%%%%%%%%%%%%%%%%%%%%%%%%%%%%%%%%%%%%%%%%%%%%%%%%%
\section{Introduction}
\label{sec1}
%%%%%%%%%%%%%%%%%%%%%%%%%%%%%%%%%%%%%%%%%%%%%%%%%%%%

I never had the pleasure of meeting Dubravko Tadi$\acute{\rm c}$ ,
which is a shame, since he and I worked on many parallel subjects
over the years.  An example of this is his recent work on
hypernuclear decay\cite{t1} as well as his early papers on what is
usually called nuclear parity violation\cite{t2}.  It is the
latter which I wish to focus on in this paper, which I dedicate to
Dubravko's memory.

The cornerstone of traditional nuclear physics is the study of
nuclear forces and, over the years, phenomenological forms of the
nuclear potential have become increasingly sophisticated. In the
nucleon-nucleon ($NN$) system, where data abound, the present
state of the art is indicated, for example, by phenomenological
potentials such as AV18 that are able to fit phase shifts in the
energy region from threshold to 350 MeV in terms of $\sim$ 40
parameters. Progress has also been made in the description of
few-nucleon systems \cite{few}.  At the same time, in recent years
a new technique ---effective field theory (EFT)--- has been used
in order to attack this problem using the symmetries of QCD
\cite{eft}. In this approach the nuclear interaction is separated
into long- and short-distance components. In its original
formulation \cite{wei}, designed for processes with typical
momenta comparable to the pion mass, $Q\sim m_\pi$, the
long-distance component is described fully quantum mechanically in
terms of pion exchange, while the short-distance piece is
described in terms of a small number of
phenomenologically-determined contact couplings. The resulting
potential \cite{ray,3Npot} is approaching \cite{fits,chiralfew}
the degree of accuracy of purely-phenomenological potentials. Even
higher precision can be achieved at lower momenta, where all
interations can be taken as short-ranged, as has been demonstrated
not only in the $NN$ system \cite{aleph,crs}, but also in the
three-nucleon system \cite{3stooges,triton}. Precise
---$\sim 1\%$--- values have been generated also for low-energy,
astrophysically-important cross sections for reactions such as
$n+p\rightarrow d+\gamma$ \cite{npd} and $p+p\rightarrow
d+e^++\nu_e$\cite{sun}. However, besides providing reliable values
for such quantities, the use of EFT techniques allows for the a
realistic estimation of the size of possible corrections.

Over the past nearly half century there has also developed a
series of measurements attempting to illuminate the parity-{\it
violating} (PV) nuclear interaction.  Indeed the first
experimental paper of which I am aware was that of Tanner in 1957
\cite{tan}, shortly after the experimental confirmation of parity
violation in nuclear beta decay by Wu et al. \cite{par}. Following
seminal theoretical work by Michel in 1964 \cite{mic} and that of
other authors in the late 1960's \cite{ope,oth,pir}, the results
of such experiments have generally been analyzed in terms of a
meson-exchange picture, and in 1980 the work of Desplanques,
Donoghue, and Holstein (DDH) developed a comprehensive and general
meson-exchange framework for the analysis of such interactions in
terms of seven parameters representing weak parity-violating
meson-nucleon couplings \cite{ddh}. The DDH interaction has become
the standard setting by which hadronic and nuclear PV processes
are now analyzed theoretically.

It is important to observe, however, that the DDH framework is, at
heart, a {\em model} based on a meson-exchange picture. Provided
one is interested primarily in near-threshold phenomena, use of a
model is unnecessary, and one can instead represent the PV nuclear
interaction in a model-independent effective-field-theoretic
fashion, as recently developed by Zhu et al.\cite{zhu}.  In this
approach, the low energy PV NN interaction is entirely
short-ranged, and the most general potential depends at leading
order on 11 independent operators parameterized by a set of 11
{\em a priori} unknown low-energy constants (LEC's). When applied
to low-energy ($E_{\rm cm}\leq 50$ MeV) two-nucleon PV
observables, however, such as the neutron spin asymmetry in the
capture reaction ${\vec n}+p\to d+\gamma$, the 11 operators reduce
to a set of five independent PV amplitudes which may be determined
by an appropriate set of measurements, as described in \cite{zhu},
and an experimental program which should result in the
determination of these couplings is underway.  This is an
important goal, since such interactions are interesting not only
in their own right but also as background effects entering atomic
PV measurements\cite{ana} as well as experiments that use parity
violation in electromagnetic interactions in order to probe
nuclear structure\cite{str1}.

Completion of such a low-energy program would serve at least three
additional purposes:

\begin{itemize}

\item [i)] First, it would provide particle theorists with a set of five
benchmark numbers which are in principle explainable from first
principles. This situation would be analogous to what one
encounters in chiral perturbation theory for pseudoscalars, where
the experimental determination of the ten LEC's appearing in the
${\cal O}(p^4)$ Lagrangian presents a challenge to hadron
structure theory. While many of the ${\cal O}(p^4)$ LEC's are
saturated by $t$-channel exchange of vector mesons, it is not
clear {\em a priori} that the analogous PV NN constants are
similarly saturated (as assumed implicitly in the DDH model).

\item [ii)] Moreover, analysis of the PV NN LEC's involves the interplay
of weak and strong interactions in the strangeness conserving
sector. A similar situation occurs in $\Delta S=1$ hadronic weak
interactions, and the interplay of strong and weak interactions in
this case are both subtle and only partially understood, as
evidenced, {\em e.g.}, by the well-known the $\Delta I=1/2$ rule
enigma. The additional information in the $\Delta S=0$ sector
provided by a well-defined set of experimental numbers would
undoubtedly shed light on this fundamental problem.

\item [iii)] Finally, the information derived from the low-energy nuclear PV program
would also provide a starting point for a reanalysis of PV effects
in many-body systems. Until now, one has attempted to use PV
observables obtained from both few- and many-body systems in order
to determine the seven PV meson-nucleon couplings entering the DDH
potential, and several inconsistencies have emerged. The most
blatant is the vastly different value for $h_\pi$ obtained from
the PV $\gamma$-decays of $^{18}$F, $^{19}$F and from the
combination of the ${\vec p}p$ asymmetry and the cesium anapole
moment\cite{ana}. The origin of this clash could be due to any one
of a number of factors. Using the operator constraints derived
from the few-body program as input into the nuclear analysis could
help clarify the situation. It may be, for example, that the
remaining combinations of operators not constrained by the
few-body program play a more significant role in nuclei than
implicitly assumed by the DDH framework. Alternatively, truncation
of the model space in shell model treatments of the cesium anapole
moment may be the culprit. In any case, approaching the nuclear
problem from a more systematic perspective and drawing upon the
results of few-body studies would undoubtedly represent an advance
for the field.

\end{itemize}

The purpose of the present paper is not, however, to make the case
for the effective field theory program---this has already been
undertaken in \cite{zhu}.  Also, it is not our purpose to review
the subject of hadronic parity violation---indeed there exist a
number of comprehensive recent reviews of this
subject\cite{hhr}\cite{des}\cite{hax}. However, although the basic
ideas of the physics are clearly set out in these works, because
the NN interaction is generally represented in terms of a somewhat
forbidding two-body interaction, any calculations which are done
involve state of the art potentials and are somewhat mysterious
except to those priests who preach this art. Rather, in this
paper, we wish to argue that this need not be the case. Below we
eschew a high precision but complex nuclear wavefunction approach
in favor of a simple analytic treatment which captures the flavor
of the subject without the complications associated with a more
rigorous calculation. We show that, provided that one is working
in the low energy region, one can use a simple effective
interaction approach to the PV NN interaction wherein the the
parity violating NN interaction is described in terms of just five
real numbers, which characterize S-P wave mixing in the spin
singlet and triplet channels, and the experimental and theoretical
implications can be extracted within a basic effective interaction
technique, wherein the nucleon interactions are represented by
short range potentials. This is justified at low energy because
the scales indicated by the scattering lengths---$a_s\sim-20$ fm,
$a_t\sim 5$ fm---are both much larger than the $\sim 1$ fm range
of the nucleon-nucleon strong interaction.   Of course, precision
analysis should still be done with the best and most powerful
contemporary wavefunctions such as the Argonne V18 or Bonn
potentials.  Nevertheless, for a simple introduction to the field,
we feel that the elementary discussion given below is didactically
and motivationally useful. In the next section then we present a
brief review of the standard DDH formalism, since this is the
basis of most analysis, as well as the EFT picture in which we
shall work. Then in the following section we show how the basic
physics of the NN system can be elicited in a simple analytic
fashion, focusing in particular on the deuteron. With this as a
basis we proceed to the parity-violating NN interaction and
develop a simple analytic description of low energy PV processes.
We summarize out findings in a brief concluding section

%%%%%%%%%%%%%%%%%%%%%%%%%%%%%%%%%%%%%%%%%%%%%%%%%%%%%%%%%%%%%%%%%%%%%%%%%%%%%%%

\section{Hadronic Parity Violation: Old and New}
\label{sec2a}
%%%%%%%%%%%%%%%%%%%%%%%%%%%%%%%%%%%%%%%%%%%%%%%%%%%%%%%%%%%%%%%%%%%%%%%%%%%%%%%

The essential idea behind the conventional DDH framework relies on
the fairly successful representation of the parity-conserving $NN$
interaction in terms of a single meson-exchange approach.  Of
course, this technique requires the use of strong interaction
couplings of the lightest vector and pseudoscalar mesons
\begin{eqnarray}
{\cal H}_{\rm st}&=& ig_{\pi NN}\bar{N}\gamma_5\tau\cdot\pi N
+g_\rho\bar{N}\left(\gamma_\mu
                    +i{\chi_\rho\over 2m_N}\sigma_{\mu\nu}k^\nu\right)
\tau\cdot\rho^\mu N\nonumber\\
&&+g_\omega\bar{N}\left(\gamma_\mu
                        +i{\chi_\omega\over 2m_N}\sigma_{\mu\nu}k^\nu
\right)\omega^\mu N, \label{eq:pch}
\end{eqnarray}
whose values are reasonably well determined. The DDH approach to
the parity-violating weak interaction utilizes a similar
meson-exchange picture, but now with one strong and one weak
vertex
---{\it cf.} Fig. 1.

\begin{figure}
\begin{center}
\epsfig{file=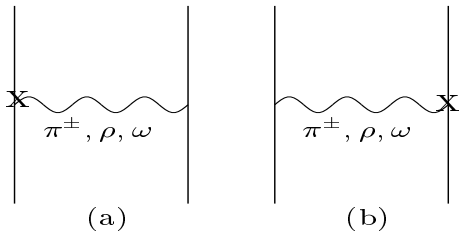,height=3cm,width=6cm}
\caption{Parity-violating $NN$ potential generated by meson
exchange.}
\end{center}
\label{fig:DDH}
\end{figure}

We require then an effective parity-violating $NNM$ Hamiltonian in
analogy to Eq. (\ref{eq:pch}).  The process is simplified somewhat
by Barton's theorem, which requires that, in the CP-conserving
limit, which we employ, exchange of neutral pseudoscalars is
forbidden \cite{bar}.  From general arguments, the effective
Hamiltonian for such interactions must take the form
\begin{eqnarray}
{\cal H}_{\rm wk} &=&i{h_\pi\over
\sqrt{2}}\bar{N}(\tau\times\pi)_3N
+\bar{N}\left(h_\rho^0\tau\cdot\rho^\mu +h_\rho^1\rho_3^\mu
+{h_\rho^2\over 2\sqrt{6}}(3\tau_3\rho_3^\mu
-\tau\cdot\rho^\mu)\right)
\gamma_\mu\gamma_5N\nonumber\\
&&+\bar{N}
\left(h_\omega^0\omega^\mu+h_\omega^1\tau_3\omega^\mu\right)\gamma_\mu\gamma_5N
-h_\rho^{'1}\bar{N}(\tau\times\rho^\mu)_3{\sigma_{\mu\nu}k^\nu\over
2m_N} \gamma_5N.
\end{eqnarray}
We see that there exist, in this model, seven unknown weak
couplings $h_\pi$, $h_\rho^0$, ...  However, quark model
calculations suggest that $h_\rho^{'1}$ is quite small \cite{bhh},
so this term is usually omitted, leaving parity-violating
observables described in terms of just six constants.  DDH
attempted to evaluate such PV couplings using basic quark-model
and symmetry techniques, but they encountered significant
theoretical uncertainties.  For this reason their results were
presented in terms of an allowable range for each, accompanied by
a ``best value'' representing their best guess for each coupling.
These ranges and best values are listed in Table \ref{tab0},
together with predictions generated by subsequent groups
\cite{dz,fcdh}.

\begin{table}
\begin{center}
\begin{tabular}{|c|c|c|c|c|}
\hline \quad   & DDH\cite{ddh} & DDH\cite{ddh} & DZ\cite{dz} &
FCDH\cite{fcdh}\\
Coupling & Reasonable Range & ``Best" Value &  &  \\ \hline
$f_\pi$ & $0\rightarrow 30$ &+12&+3&+7\\
$h_\rho^0$& $30\rightarrow -81$&$-30$&$-22$&$-10$\\
$h_\rho^1$& $-1\rightarrow 0$& $-0.5$&+1&$-1$\\
$h_\rho^2$& $-20\rightarrow -29$&$-25$&$-18$&$-18$\\
$h_\omega^0$&$15\rightarrow -27$&$-5$&$-10$&$-13$\\
$h_\omega^1$&$-5\rightarrow -2$&$-3$&$-6$&$-6$\\ \hline
\end{tabular}
\caption{Weak $NNM$ couplings as calculated in Refs.
\cite{ddh,dz,fcdh}. All numbers are quoted in units of the ``sum
rule" value $g_\pi =3.8\cdot 10^{-8}$.}
\end{center}
\label{tab0}
\end{table}

Before making contact with experimental results, however, it is
necessary to convert the $NNM$ couplings generated above into an
effective parity-violating $NN$ potential.  Inserting the strong
and weak couplings, defined above into the meson-exchange diagrams
shown in Fig. 1 and taking the Fourier transform, one finds the
DDH effective parity-violating $NN$ potential
\begin{eqnarray}
V^{\rm PV}_{DDH}(\vec{r}) &=&i{h_\pi g_{\pi NN}\over
\sqrt{2}}\left({\tau_1\times\tau_2\over 2}\right)_3
(\vec{\sigma}_1+\vec{\sigma}_2)\cdot
\left[{{\vec p}_1-{\vec p}_2\over 2m_N},w_\pi (r)\right]\nonumber\\
&&-g_\rho\left(h_\rho^0\tau_1\cdot\tau_2+h_\rho^1\left({\tau_1+\tau_2\over
2} \right)_3+h_\rho^2{(3\tau_1^ 3\tau_2^3-\tau_1\cdot\tau_2)\over
2\sqrt{6}}\right)
\nonumber\\
&&\quad \left((\vec{\sigma}_1-\vec{\sigma}_2)\cdot \left\{{{\vec
p}_1-{\vec p}_2\over 2m_N},w_\rho(r)\right\}
+i(1+\chi_V)\vec{\sigma}_1\times\vec{\sigma}_2\cdot \left[{{\vec
p}_1-{\vec p}_2\over 2m_N},w_\rho
(r)\right]\right)\nonumber\\
&&-g_\omega\left(h_\omega^0+h_\omega^1\left({\tau_1+\tau_2\over
2}\right)_3
\right)\nonumber\\
&&\quad\left((\vec{\sigma}_1-\vec{\sigma}_2)\cdot \left\{{{\vec
p}_1-{\vec p}_2\over 2m_N},w_\omega (r)\right\}
+i(1+\chi_S)\vec{\sigma}_1\times\vec{\sigma}_2\cdot \left[{{\vec
p}_1-{\vec p}_2\over
2m_N},w_\omega(r)\right]\right)\nonumber\\
&&-\left(g_\omega h_\omega^1-g_\rho h_\rho^1\right)
\left({\tau_1-\tau_2\over 2}\right)_3
(\vec{\sigma}_1+\vec{\sigma}_2)\cdot \left\{{{\vec p}_1-{\vec
p}_2\over 2m_N},w_\rho(r)\right\}
\nonumber\\
&&-g_\rho h_\rho^{1'}i\left({\tau_1\times\tau_2\over 2}\right)_3
(\vec{\sigma}_1+\vec{\sigma}_2)\cdot \left[{{\vec p}_1-{\vec
p}_2\over 2m_N},w_\rho(r)\right],
\end{eqnarray}
where $w_i(r)=\exp (-m_ir)/4\pi r$ is the usual Yukawa form,
$r=|{\vec x}_1 - {\vec x}_2|$ is the separation between the two
nucleons, and ${\vec p}_i=-i{\vec\nabla}_i$.

Nearly all experimental results involving nuclear parity violation
have been analyzed using $V^{\rm PV}_{DDH}$ for the past
twenty-some years. At present, however, there appear to exist
discrepancies between the values extracted for the various DDH
couplings from experiment. In particular, the values of $h_\pi$
and $h_\rho^0$ extracted from $\vec{p}p$ scattering and the
$\gamma$ decay of $^{18}$F do not appear to agree with the
corresponding values implied by the anapole moment of $^{133}$Cs
measured in atomic parity violation \cite{atomic}.

These inconsistencies suggest that the DDH framework may not,
after all, adequately characterize the PV NN interaction and
provides motivation for our reformulation using EFT.  In this
approach, the effective PV potential is entirely short-ranged and
has the co-ordinate space form
\begin{eqnarray}\label{3}\nonumber
V_{eff}^{PV} ({\vec r}) &=& {2\over \Lambda_\chi^3} \left\{
 \left[ C_1 + C_2 {\tau_1^z +\tau_2^z\over 2}\right]
\left( {\vec \sigma}_1 -{\vec \sigma}_2\right)\cdot \{-i\vec{\nabla},f_m(r)\} \right.\\
\nonumber &&\left. +  \left[ {\tilde C}_1  +  {\tilde C}_2
{\tau_1^z +\tau_2^z\over 2} \right] i \left( {\vec \sigma}_1
\times {\vec \sigma}_2 \right) \cdot [-i\vec{\nabla},f_m(r)]
\right.\\ \nonumber &&\left. + \left[ C_2 -C_4 \right] {\tau_1^z
-\tau_2^z\over 2}
\left( {\vec \sigma}_1 +{\vec \sigma}_2\right) \cdot \{-i\vec{\nabla},f_m(r)\}  \right.\\
\nonumber &&\left. + \left[ C_3 \tau_1 \cdot \tau_2 + C_4
{\tau_1^z +\tau_2^z\over 2} +{\cal I}_{ab} C_5 \tau_1^a \tau_2^b
\right] \left( {\vec \sigma}_1 -{\vec \sigma}_2\right)\cdot
\{-i\vec{\nabla},f_m(r)\} \right.\\ \nonumber &&\left. + \left[
{\tilde C}_3 \tau_1 \cdot \tau_2 + {\tilde C}_4{\tau_1^z
+\tau_2^z\over 2} +{\cal I}_{ab} {\tilde C}_5\tau_1^a \tau_2^b
\right] i \left( {\vec \sigma}_1 \times {\vec \sigma}_2 \right)
\cdot [-i\vec{\nabla},f_m(r)]
\right.\\
&&\left. + C_6 i\epsilon^{ab3} \tau_1^a \tau_2^b \left( {\vec
\sigma}_1 +{\vec \sigma}_2\right)\cdot [-i\vec{\nabla},f_m(r)]
\right\}, \label{eq:sht}
\end{eqnarray}
where
\begin{equation}
{\cal I}^{ab}= \left(\begin{array}{lll}1&0&0\\0&1&0\\0&0&-2
 \end{array} \right),\label{eq:chg}
\end{equation}
and $f_m({\vec r})$ is a function which \begin{itemize}
\item [i)]
is strongly peaked, with width $\sim 1/m$ about $r=0$, and
\item [ii)] approaches
$\delta^{(3)}({\vec r})$ in the zero width---$m\rightarrow
\infty$---limit.
\end{itemize}
A convenient form, for example, is the Yukawa-like function
\begin{equation}
f_m(r)={m^2\over 4\pi r}\exp (-mr)\label{eq:yuk}
\end{equation}
where $m$ is a mass chosen to reproduce the appropriate short
range effects. Actually, for the purpose of carrying out actual
calculations, one could just as easily use the momentum-space form
of $V^{\rm PV}_{\rm SR}$, thereby avoiding the use of $f_m({\vec
r})$ altogether. Nevertheless, the form of Eq. \ref{eq:sht} is
useful when comparing with the DDH potential. For example, we
observe that the same set of spin-space and isospin structures
appear in both $V^{\rm PV}_{eff}$ and the vector-meson exchange
terms in $V^{\rm PV}_{\rm DDH}$, though the relationship between
the various coefficients in $V^{\rm PV}_{eff}$ is more general. In
particular, the DDH model is tantamount to assuming
\begin{equation}
{ {\tilde C}_1 \over C_1}= { {\tilde C}_2 \over C_2}=
1+\chi_\omega,
\end{equation}
\begin{equation}
{ {\tilde C}_3 \over C_3}= { {\tilde C}_4 \over C_4}={ {\tilde
C}_5 \over C_5}=
 1+\chi_\rho,\label{eq:ddhr}
\end{equation}
and taking $m\sim m_\rho,m_\omega$, assumptions which may not be
physically realistic.  Nevertheless, if this ansatz is made, the
EFT and DDH results coincide provided the identifications

\begin{eqnarray}\nonumber
C_1^{DDH} &=&-{ \Lambda_\chi^3 \over 2 m_N m_\omega^2} g_{\omega
NN} h_\omega^0,\\ \nonumber
 C_2^{DDH}& =&-{ \Lambda_\chi^3\over 2 m_N m_\omega^2} g_{\omega NN}
h_\omega^1,\\ \nonumber
 C_3^{DDH} &=&-{\Lambda_\chi^3\over 2 m_N m_\rho^2}g_{\rho NN}
h_\rho^0,  \\
\nonumber
 C_4^{DDH} &=&-{\Lambda_\chi^3 \over 2 m_N m_\rho^2}g_{\rho NN}
h_\rho^1, \\
\nonumber
 C_5^{DDH} &=&{ \Lambda_\chi^3\over 4\sqrt{6} m_N m_\rho^2}g_{\rho NN}
h_\rho^2, \\ \nonumber
 C_6^{DDH}&=& -{\Lambda_\chi^3 \over 2 m_N m_\rho^2} g_{\rho NN}
h_\rho^{\prime 1}.\label{eq:rel}
\end{eqnarray}
are made\cite{zhu}.

Before beginning our analysis of PV NN scattering, however, it is
important to review the analogous PC NN scattering case, since it
is more familiar and it is a useful arena wherein to compare
conventional and effective field theoretic methods.

\section{Parity Conserving NN Scattering}

We begin our discussion with a brief review of conventional
scattering theory\cite{merz}. In the usual partial wave expansion,
we can write the scattering amplitude as
\begin{equation}
f(\theta)=\sum_\ell(2\ell+1)a_\ell(k)P_\ell(\cos\theta)
\end{equation}
where $a_\ell(k)$ has the form
\begin{equation}
a_\ell(k)={1\over k}e^{i\delta(k)}\sin\delta(k)={1\over
k\cot\delta(k) -ik}\label{eq:zz}
\end{equation}
\subsection{Conventional Analysis}
Working in the usual potential model approach, a general
expression for the scattering phase shift $\delta_\ell(k)$
is\cite{merz}
\begin{equation}
\sin\delta_\ell(k)=-k\int_0^\infty
dr'r'j_\ell(kr')2m_rV(r')u_{\ell,k}(r')\label{eq:sl}
\end{equation}
where $m_r$ is the reduced mass and
\begin{eqnarray}
u_{\ell,k}(r)&=&r\cos\delta_\ell(k)j_\ell(kr)
+kr\int_0^rdr'r'j_\ell(kr')n_\ell(kr)u_{\ell,k}(r')2m_rV(r')\nonumber\\
&+&kr\int_r^\infty
dr'r'j_\ell(kr)n_\ell(kr')u_{\ell,k}(r')2m_rV(r')
\end{eqnarray}
is the scattering wavefunction.  At low energies one can
characterize the analytic function $k^{2\ell+1}\cot\delta(k)$ via
an effective range expansion\cite{part}
\begin{equation}
k^{2\ell+1}\cot\delta_\ell(k)=-{1\over a}+{1\over
2}r_ek^2+\ldots\label{eq:er}
\end{equation}
Then from Eq. \ref{eq:sl} we can identify the scattering length as
\begin{equation}
a_\ell={1\over [(2\ell+1)!!]^2}\int_0^\infty
dr'(r')^{2\ell+2}2m_rV(r') +{\cal O}(V^2)\label{eq:a}
\end{equation}
For simplicity, we consider only S-wave interactions. Then for
neutron-proton interactions, for example, one finds
\begin{eqnarray}
a_0^s&=&-23.715\pm 0.015\,\,{\rm fm},\,\,r_0^s=2.73\pm
0.03\,\,{\rm fm}
\nonumber\\
a_0^t&=&5.423\pm 0.005\,\,{\rm fm},\,\,r_0^t=1.73\pm 0.02\,\,{\rm
fm} \label{eq:st}
\end{eqnarray}
for scattering in the spin-singlet and spin-triplet channels
respectively.  The existence of a bound state $E_B=-\gamma^2/2m_r$
is indicated by the presence of a pole along the positive
imaginary $k$-axis---{\it i.e.} $\gamma>0$ under the analytic
continuation $k\rightarrow i\gamma$---
\begin{equation}
 {1\over a_0}+{1\over 2}r_0\gamma^2-\gamma=0
\end{equation}
We see from Eq. \ref{eq:st} that there is no bound state in the np
spin-singlet channel, but in the spin-triplet system there exists
a solution
\begin{equation}
\kappa={1-\sqrt{1-{2r_0^t\over a_0^t}}\over r_0^t}=45.7\,\,{\rm
MeV},\quad i.e.\quad E_B=-2.23\,\, {\rm MeV}
\end{equation}
corresponding to the deuteron.

As a specific example, suppose we utilize a simple square well
potential to describe the interaction
\begin{equation}
V(r)=\left\{\begin{array}{cc}
-V_0 & r\leq R \\
0 &  r>R
\end{array}\right.
\end{equation}
For S-wave scattering the wavefunction in the interior and
exterior regions can then be written as
\begin{equation}
\psi^{(+)}(r)=\left\{\begin{array}{lc}
Nj_0(Kr) & r\leq R \\
N'(j_0(kr)\cos\delta_0-n_0(kr)\sin\delta_0) & r>R
\end{array}\right.
\end{equation}
where $j_0,n_0$ are spherical harmonics and the interior, exterior
wavenumbers are given by $k=\sqrt{2m_rE}$, $K=\sqrt{2m_r(E+V_0)}$
respectively.  The connection between the two forms can be made by
matching logarithmic derivatives at the boundary, which yields
\begin{equation}
k\cot\delta\simeq -{1\over R}\left[1+{1\over KRF(KR)}\right]\quad
{\rm with} \quad F(x)=\cot x-{1\over x}
\end{equation}
Making the effective range expansion---Eq \ref{eq:er}---we find an
expression for the scattering length
\begin{equation}
a_0= R\left[1-{\tan(K_0R)\over K_0R}\right]\quad {\rm where}\quad
K_0=\sqrt{2m_rV_0}
\end{equation}
Note that for weak potentials---$K_0R<<1$---this form agrees with
the general result Eq. \ref{eq:a}---
\begin{equation}
a_0=\int_0^\infty dr'{r'}^22m_rV(r')= -{2m_r\over 3} R^3V_0+{\cal
O}(V_0^2)\label{eq:cc}
\end{equation}

\subsection{Coulomb Effects}

When Coulomb interactions are included the analysis becomes
somewhat more challenging. Suppose first that only same charge
({\it e.g.}, proton-proton) scattering is considered and that, for
simplicity, we describe the interaction in terms of a potential of
the form
\begin{equation}
V(r)=\left\{\begin{array}{cc}
U(r)& r< R\\
\alpha\over r& r>R
\end{array}
\right.
\end{equation}
{\it i.e.} a strong attraction---$U(r)$---at short distances, in
order to mimic the strong interaction, and the repulsive Coulomb
potential---$\alpha/r$---at large distance, where $\alpha\simeq
1/137$ is the fine structure constant. The analysis of the
scattering then proceeds as above but with the replacement of the
exterior spherical Bessel functions by corresponding Coulomb
wavefunctions $F_0^+,G_0^+$
\begin{equation}
j_0(kr)\rightarrow F_0^+(r),\qquad n_0(kr)\rightarrow G_0^+(r)
\end{equation}
whose explicit form can be found in reference \cite{bj}.  For our
purposes we require only the form of these functions in the limit
$kr<<1$---
\begin{eqnarray}
F_0^+(r)&\stackrel{kr<<1}{\longrightarrow}&C(\eta_+(k))(1+
{r\over 2a_B}+\ldots)\nonumber\\
G_0^+(r)&\stackrel{kr<<1}{\longrightarrow}&-{1\over
C(\eta_+(k))}\left\{{1\over
kr}\right.\nonumber\\
&+&\left.2\eta_+(k)\left[h(\eta_+(k))+2\gamma_E-1+\ln {r\over
a_B}\right]+\ldots\right\}\nonumber\\
\quad\label{eq:tl}
\end{eqnarray}
Here $\gamma_E=0.577215..$ is the Euler constant,
\begin{equation}
C^2(x)={2\pi x\over \exp(2\pi x)-1}
\end{equation}
is the usual Coulombic enhancement factor, $a_B=1/m_r\alpha$ is
the Bohr radius, $\eta_+(k)=1/2ka_B$, and
\begin{equation}
h(\eta_+(k))={\rm Re}H(i\eta_+(k))= \eta_+^2(k) \sum_{n=1}^\infty
{1\over n(n^2+\eta_+^2(k))}-\ln\eta_+(k)-\gamma_E
\end{equation}
where $H(x)$ is the analytic function
\begin{equation}
H(x)=\psi(x)+{1\over 2x}-\ln(x)
\end{equation}
Equating interior and exterior logarithmic derivatives we find
\begin{eqnarray}
KF(KR)&=&{\cos\delta_0 {F_0^+}'(R)-\sin\delta_0{G_0^+}'(R)\over
\cos\delta_0F_0^+(R)
-\sin\delta_0G_0^+(R)}\nonumber\\
&=&{k\cot\delta_0C^2(\eta_+(k)){1\over 2a_B}-{1\over R^2}\over
k\cot\delta_0C^2(\eta_+(k))+{1\over R}+{1\over
a_B}\left[h(\eta_+(k))-\ln{a_B\over R}+2\gamma_E-1\right]}\nonumber\\
\quad\label{eq:aa}
\end{eqnarray}
Since $R<<a_B$ Eq. \ref{eq:aa} can be written in the form
\begin{equation}
k\cot\delta_0C^2(\eta_+(k))+{1\over
a_B}\left[h(\eta_+(k))-\ln{a_B\over R}+2\gamma_E-1\right]\simeq
-{1\over a_0}
\end{equation}
The scattering length $a_C$ in the presence of the Coulomb
interaction is conventionally defined as\cite{pres}
\begin{equation}
k\cot\delta_0C^2(\eta_+(k))+{1\over a_B}h(\eta_+(k))=-{1\over a_C}
+\ldots\label{eq:dd}
\end{equation}
so that we have the relation
\begin{equation}
-{1\over a_0}=-{1\over a_C}-{1\over a_B}(\ln{a_B\over
R}+1-2\gamma_E)\label{eq:yy}
\end{equation}
between the experimental scattering length---$a_C$---and that
which would exist in the absence of the Coulomb
interaction---$a_0$.

As an aside we note that, strictly speaking, $a_0$ is not itself
an observable since the Coulomb interaction {\it cannot} be turned
off. However, in the case of the pp interaction isospin invariance
requires $a_0^{pp}=a_0^{nn}$ so that one has the prediction
\begin{equation}
-{1\over a_0^{nn}}=-{1\over a_C^{pp}}-\alpha M_N(\ln{1\over \alpha
M_NR}+1 -2\gamma_E)\label{eq:oo}
\end{equation}
While this is a model dependent result, Jackson and Blatt have
shown, by treating the interior Coulomb interaction
perturbatively, that a version of this result with
$1-2\gamma_E\rightarrow 0.824-2\gamma_E$ is approximately valid
for a wide range of strong interaction potentials\cite{bj} and the
correction indicated in Eq. \ref{eq:oo} is essential in restoring
agreement between the widely discrepant---$a_0^{nn}=-18.8$ fm vs.
$a_C^{pp}=-7.82$ fm---values obtained experimentally.

Returning to the problem at hand, the experimental scattering
amplitude can then be written as
\begin{eqnarray}
f_C^+(k)&=&{e^{2i\sigma_0}C^2(\eta_+(k))\over -{1\over
a_C}-{1\over
a_B}h(\eta_+(k))-ikC^2(\eta_+(k))}\nonumber\\
&=&{e^{2i\sigma_0}C^2(\eta_+(k))\over -{1\over a_C}-{1\over
a_B}H(i\eta_+(k))}\label{eq:hh}
\end{eqnarray}
where $\sigma_0={\rm arg}\Gamma(1-i\eta_+(k))$ is the Coulomb
phase.

\subsection{Effective Field Theory Analysis}

Identical results may be obtained using effective field theory
(EFT) methods and in many ways the derivation is clearer and more
intuitive\cite{kap}. The basic idea here is that since we are only
interested in interactions at very low energy, a scattering length
description is quite adequate. From Eq. \ref{eq:cc} we see that,
at least for weak potentials, the scattering length has a natural
representation in terms of the momentum space potential
$\tilde{V}(\vec{p}=0)$---
\begin{equation}
a_0={m_r\over 2\pi}\int d^3rV(r)={m_r\over
2\pi}\tilde{V}(\vec{p}=0)
\end{equation}
and it is thus natural to perform our analysis using a simple
contact interation.  First consider the situation that we have two
particles A,B interacting only via a local strong interaction, so
that the effective Lagrangian can be written as
\begin{equation}
{\cal L}=\sum_{i=A}^B\Psi_i^\dagger(i{\partial\over \partial
t}+{\vec{\nabla}^2\over
2m_i})\Psi_i-C_0\Psi_A^\dagger\Psi_A\Psi_B^\dagger\Psi_B
+\ldots\label{eq:gg}
\end{equation}
The T-matrix is then given in terms of the multiple scattering
series shown in Figure 1
\begin{equation}
T_{fi}(k)=-{2\pi\over
m_r}f(k)=C_0+C_0^2G_0(k)+C_0^3G_0^2(k)+\ldots ={C_0\over
1-C_0G_0(k)}\label{eq:ll}
\end{equation}
where $G_0(k)$ is the amplitude for particles $A,B$ to travel from
zero separation to zero separation---{\it i.e} the propagator
$D_F(k;\vec{r}'=0,\vec{r}=0)$---
\begin{equation}
G_0(k)=\lim_{\vec{r}',\vec{r}\rightarrow 0}\int{d^3s\over
(2\pi)^3}{ e^{i\vec{s}\cdot\vec{r}'}e^{-i\vec{s}\cdot\vec{r}}\over
{k^2\over 2m_r}-{s^2\over 2m_r}+i\epsilon}=\int{d^3s\over
(2\pi)^3} {2m_r\over k^2-s^2+i\epsilon}
\end{equation}
Equivalently $T_{fi}(k)$ satisfies a Lippman-Schwinger equation
\begin{equation}
T_{fi}(k)=C_0+C_0G_0(k)T_{fi}(k).
\end{equation}
whose solution is given in Eq. \ref{eq:ll}.

\begin{figure}[htb]
\begin{center}
\epsfig{figure=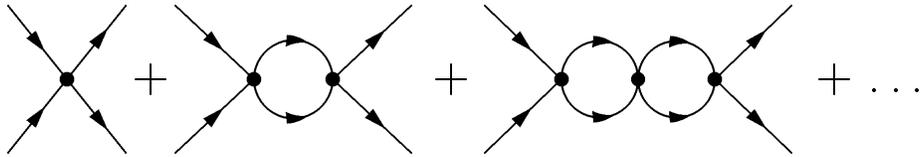,height=20mm}
\end{center}
\vspace{-8mm} \caption{The multiple scattering series.}
\label{fig}
\end{figure}

The complication here is that the function $G_0(k)$ is divergent
and must be defined via some sort of regularization.  There are a
number of ways by which to do this, but perhaps the simplest is to
use a cutoff regularization with $k_{max}=\mu$, which simply
eliminates the high momentum components of the wavefunction
completely.  Then
\begin{equation}
G_0(k)=-{m_r\over 2\pi}({2\mu\over \pi}+ik)
\end{equation}
(Other regularization schemes are similar.  For example, one could
subtract at an unphysical momentum point, as proposed by
Gegelia\cite{geg}
\begin{equation}
G_0(k)=\int{d^3s\over (2\pi)^3}({2m_r\over
k^2-s^2+i\epsilon}+{2m_r\over \mu^2+s^2})=-{m_r\over 2\pi}(\mu+ik)
\end{equation}
which has been shown by Mehen and Stewart\cite{ms} to be
equivalent to the power divergence subtraction (PDS) scheme
proposed by Kaplan, Savage and Wise.\cite{kap}) In any case, the
would-be linear divergence is, of course, cancelled by
introduction of a counterterm accounting for the omitted high
energy component of the theory, which renormalizes $C_0$ to
$C_0(\mu)$. That $C_0(\mu)$ should be a function of the cutoff is
clear because by varying the cutoff energy we are varying the
amount of higher energy physics which we are including in our
effective description. The scattering amplitude then becomes
\begin{equation}
f(k)=-{m_r\over 2\pi}\left({1\over{1\over
C_0(\mu)}-G_0(k)}\right)={1\over -{2\pi\over
m_rC_0(\mu)}-{2\mu\over \pi}-ik}
\end{equation}
Comparing with Eq. \ref{eq:zz} we identify the scattering length
as
\begin{equation}
-{1\over a_0}=-{2\pi\over m_rC_0(\mu)}-{2\mu\over \pi}
\end{equation}
Of course, since $a_0$ is a physical observable, it is cutoff
independent, so that the $\mu$ dependence of $1/C_0(\mu)$ is
cancelled by the cutoff dependence in the Green's function.

\subsection{Coulomb Effects in EFT}

More interesting is the case where we restore the Coulomb
interaction between the particles.  The derivatives in Eq.
\ref{eq:gg} then become covariant and the bubble sum is evaluated
with static photon exchanges between each of the lines---each
bubble is replaced by one involving a sum of zero, one, two, etc.
Coulomb interactions, as shown in Figure 2.

\begin{figure}[htb]
\begin{center}
\epsfig{figure=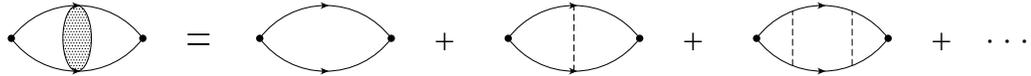,height=10mm}
\end{center}
\vspace{-5mm} \caption{The Coulomb corrected bubble.} \label{fig2}
\end{figure}

The net result in the case of same charge scattering is the
replacement of the free propagator by its Coulomb analog
\begin{eqnarray}
G_0(k)\rightarrow G_C^+(k)&=&\lim_{\vec{r}',\vec{r}\rightarrow 0}
\int{d^3s\over
(2\pi)^3}{\psi^+_{\vec{s}}(\vec{r}'){\psi^+_{\vec{s}}}^*
(\vec{r})\over {k^2\over 2m_r}-{s^2\over 2m_r}+i\epsilon}\nonumber\\
&=&\int{d^3s\over (2\pi)^3}{2m_rC^2(\eta_+(s)) \over
k^2-s^2+i\epsilon}\label{eq:uuu}
\end{eqnarray}
where
\begin{equation}
\psi^+_{\vec{s}}(\vec{r})=C(\eta_+(s))e^{i\sigma_0}
e^{i\vec{s}\cdot\vec{r}}
{}_1F_1({-i\eta_+(s),1,isr-i\vec{s}\cdot\vec{r}})
\end{equation}
is the outgoing Coulomb wavefunction for repulsive Coulomb
scattering.\cite{lal} Also in the initial and final states the
influence of static photon exchanges must be included to all
orders, which produces the factor
$C^2(2\pi\eta_+(k))\exp(2i\sigma_0)$.  Thus the repulsive Coulomb
scattering amplitude becomes
\begin{equation}
f_C^+(k)=-{m_r\over 2\pi}{C_0C^2(\eta_+(k))\exp 2i\sigma_0 \over
1-C_0{}G_C^+(k)}
\end{equation}
The momentum integration in Eq. \ref{eq:uuu} can be performed as
before using cutoff regularization, yielding
\begin{equation}
G_C^+(k)=-{m_r\over 2\pi}\left\{{2\mu\over \pi}+{1\over
a_B}\left[H(i\eta_+(k))-\ln{\mu a_B\over \pi}-\zeta\right]\right\}
\end{equation}
where $\zeta=\ln 2\pi -\gamma$. We have then
\begin{eqnarray}
f_C^+(k)={C^2(\eta_+(k))e^{2i\sigma_0}\over -{2\pi\over
m_rC_0(\mu)}-{2\mu\over \pi}-{1\over
a_B}\left[H(i\eta_+(k))-\ln{\mu
a_B\over \pi}-\zeta\right]}\nonumber\\
={C^2(\eta_+(k))e^{2i\sigma_0}\over -{1\over a_0}-{1\over
 a_B}\left[h(\eta_+(k)-\ln{\mu a_B\over \pi}-\zeta\right]
-ikC^2(\eta_+(k))}
\end{eqnarray}
Comparing with Eq. \ref{eq:hh} we identify the Coulomb scattering
length as
\begin{equation}
-{1\over a_C}=-{1\over a_0}+{1\over a_B}(\ln{\mu a_B\over
\pi}+\zeta)
\end{equation}
which matches nicely with Eq. \ref{eq:yy} if a reasonable cutoff
$\mu\sim m_\pi\sim 1/R$ is employed.  The scattering amplitude
then has the simple form
\begin{equation}
f_C^+(k)={C^2(\eta_+(k))e^{2i\sigma_0}\over -{1\over a_C}- {1\over
a_B}H(i\eta_+(k))}
\end{equation}
in agreement with Eq. \ref{eq:hh}.

Before moving to our ultimate goal, which is the parity violating
sector, it is useful to spend some additional time focusing on the
deuteron state, since this will be used in our forthcoming PV
analysis and provides a useful calibration of the precision of our
approach.

\section{The Deuteron}

 Fermi was fond of asking the
question ``Where's the hydrogen atom for this problem?'' meaning
what is the simple model that elucidates the basic physics of a
given system\cite{khr}?  In the case of nuclear structure, the
answer is clearly the deuteron, and it is essential to have a good
understanding of this simplest of nuclear systems at both the
qualitative and quantitative levels. The basic static properties
which we shall try to understand are indicated in Table 2.  Thus,
for example, from the feature that the deuteron carries unit spin
with positive parity, angular momentum arguments demand that it be
constructed from a combination of S- and D-wave components (a
P-wave piece is forbidden from parity considerations---more about
that later). Thus the wavefunction can be written in the form
\begin{equation}
\psi_d(\vec{r})={1\over \sqrt{4\pi}r}\left(u_d(r)+{3\over
\sqrt{8}}w_d(r) {\cal O}_{pn}\right)\chi_t
\end{equation}
where $\chi_t$ is the spin-triplet wavefunction and
$${\cal O}_{pn}=\vec{\sigma}_p\cdot\hat{r}\vec{\sigma}_n\cdot\hat{r}-{1\over 3}
\vec{\sigma}_p\cdot\vec{\sigma}_n$$ is the tensor operator.  Here
$u_d(r),w_d(r)$ represent the S-wave, D-wave components of the
deuteron wavefunction, respectively. We note that
\begin{eqnarray}
{\cal O}_{pn}|\uparrow\uparrow>&=&(\cos^2\theta-{1\over
3})|\uparrow\uparrow>+\sin^2\theta
e^{i2\phi}|\downarrow\downarrow>\nonumber\\
&+&\cos\theta\sin\theta
e^{i\phi}(|\uparrow\downarrow>+|\downarrow\uparrow>)
\end{eqnarray}
Using
\begin{equation}
\int{d\Omega\over 4\pi}\hat{r}_i\hat{r}_j ={1\over 3}\delta_{ij}
\end{equation}
we find the normalization condition
\begin{eqnarray}
1&=&<\psi_d|\psi_d>=\int_0^\infty dr\int d\Omega
\left[u_d^2(r)\right.\nonumber\\
&+&\left.{9\over 8}w_d^2(r)\left((\cos^2\theta-{1\over 3})^2
+2\cos^2\theta\sin^2\theta+\sin^4\theta\right)\right]\nonumber\\
&=&\int_0^\infty dr\left(u_d^2(r)+{9\over 8}w_d^2(r)(1-{2\over
9}+{1\over
9})\right)\nonumber\\
&=&\int_0^\infty dr(u_d^2(r)+w_d^2(r))
\end{eqnarray}
In lowest order we can neglect the D-wave component $w_d(r)$.
Then, in the region outside the range $r_0$ of the NN interaction
we must have
\begin{equation}
r>r_o\qquad \left(-{1\over M}{d^2\over dr^2}+{\gamma^2\over
M}\right)u_d(r) =0\label{eq:deu}
\end{equation}
where $\gamma=45.3$ MeV is the deuteron binding momentum defined
above.  The solution to Eq. \ref{eq:deu} is given by
\begin{equation}
r>r_0\qquad u_d(r)\sim e^{-\gamma r}\label{eq:deu1}
\end{equation}
However, since $1/\gamma\sim 4.3$ fm $>>r_0\sim$ 1 fm, it is a
reasonable lowest order assumption to assume the that the deuteron
wavefunction has the form Eq. \ref{eq:deu1} everywhere, so that we
may take
\begin{equation}
\psi_d(r)\sim\sqrt{\gamma\over 2\pi}{1\over r}e^{-\gamma r}
\end{equation}

\begin{table}
\begin{center}
\begin{tabular}{lcc}
Binding Energy& $E_B$ & 2.223 MeV\\
Spin-parity& $J^P$&$1^+$\\
Isospin&$T$&0\\
Magnetic Dipole Moment & $\mu_d$& 0.856$\mu_N$\\
Electric Quadrupole Moment&$Q_d$&0.286 efm$^2$\\
Charge Radius& $\sqrt{r_d^2}$& $\sim$ 2 fm
\end{tabular}
\caption{Static properties of the deuteron.}
\end{center}
\end{table}

Of course, we also must consider scattering states.  In this case
the asymptotic wavefunctions of the ${}^3S_1$ and ${}^1S_0$ states
must be of the form
\begin{equation}
\psi^{(+)}(r)\stackrel{r\rightarrow\infty}{\longrightarrow}
{e^{i\delta(k)}\over kr}\sin(kr+\delta(k))={\sin kr\over
kr}+{e^{ikr}\over r} t(k)
\end{equation}
with
$$t(k)={1\over k}e^{i\delta(k)}\sin\delta(k)$$
being the partial wave transition amplitude. At very low energy we
may use the simple effective range approximation defined above
\begin{equation}
k\,{\rm ctn}\delta(k)\simeq-{1\over a}
\end{equation}
to write
\begin{equation}
t(k)\simeq {1\over -{1\over a}-ik}
\end{equation}
Then we have the representation
\begin{equation}
\psi^{(+)}(r)\stackrel{r\rightarrow\infty}{\longrightarrow}
{e^{i\delta(k)}\over kr}\sin(kr+\delta(k))={\sin kr\over
kr}+{e^{ikr}\over r} \left({1\over -{1\over a}-ik}\right)
\end{equation}
and, comparing with a Green's function solution to the
Schr\"{o}dinger equation
\begin{equation}
\psi^{(+)}(\vec{r})=e^{i\vec{k}\cdot\vec{r}}-{M\over 4\pi}\int
d^3r' {e^{ik|\vec{r}-\vec{r}'|}\over
|\vec{r}-\vec{r}'|}U(\vec{r}')\psi(\vec{r}')
\end{equation}
we see that the potential at low energy can be represented via the
simple local potential
\begin{equation}
U(\vec{r})\simeq {4\pi\over M}a\delta^3(\vec{r})
\end{equation}
which is sometimes called the zero-range approximation (ZRA) and
is equivalent to the contact potential used in the EFT approach.

The relation between the scattering and bound state descriptions
can be obtained by using the feature that the deuteron
wavefunction must be orthogonal to its ${}^3S_1$ counterpart. This
condition reads
\begin{eqnarray}
0&=&\int
d^3r\psi_d^\dagger(r)\psi_t(r)=\sqrt{8\pi\gamma}\int_0^\infty
dre^{-\gamma r}\left({1\over k}\sin kr
+e^{ikr}t_t(k)\right)\nonumber\\
&=&\sqrt{8\pi\gamma}\left({1\over \gamma^2+k^2} +{1\over
\gamma-ik}t_t(k)\right)={\sqrt{8\pi\gamma}\over
\gamma-ik}\left({1\over \gamma+ik} +{1\over -{1\over
a_t}-ik}\right)
\end{eqnarray}
which requires that $\gamma=1/a_t$.  This necessity is also clear
from the already mentioned feature that the deuteron represents a
pole in $t_t(k)$ in the limit as $k\rightarrow i\gamma$---{\it
i.e.}, $-1/a_t+\gamma=0$.  Since ${1\over \gamma}\sim 4.3$ fm,
this equality holds to within 20\% or so and indicates the
precision of our approximation.  In spite of this roughness, there
is much which can be learned from this simple analytic approach.

We begin with the charge radius, which is defined via
\begin{equation}
<r_d^2>=<r_p^2>+\int d^3r {1\over 4}r^2|\psi_d(r)|^2
\end{equation}
Note here that we have included the finite size of the proton,
since it is comparable to the deuteron size and have scaled the
wavefunction contribution by a factor of four since
$\vec{r}_p={1\over 2}\vec{r}$. Performing the integration, we have
\begin{equation}
\int d^3r {1\over 4}r^2|\psi_d(r)|^2=\pi\int_0^\infty
drr^2u^2(r)={1\over 8\gamma^2}
\end{equation}
and, since $<r_p^2>\simeq 0.65$ fm$^2$ we find
\begin{equation}
\sqrt{<r_d^2>}=\sqrt{0.65+{1\over 8\gamma^2}}\,\,{\rm fm}\simeq
1.8\,\,{\rm fm}
\end{equation}
which is about 10\% too low and again indicates the roughness of
our approximation.

Now consider the magnetic moment, for which the relevant operator
is
\begin{eqnarray}
\vec{M}&=&{e\over
2M}\left(\mu_p\vec{\sigma}_p+\mu_n\vec{\sigma}_n\right)
+{e\over 2M}\vec{L}_p\nonumber\\
&=&{e\over 4M}\left(\vec{J}+\mu_V(\vec{\sigma}_p-\vec{\sigma}_n)
+(\mu_S-{1\over 2})(\vec{\sigma}_p+\vec{\sigma}_n)\right)
\end{eqnarray}
where $\vec{J}=\vec{L}+{1\over 2}(\vec{\sigma}_p+\vec{\sigma}_n)$
is the total angular momentum, $\mu_V=\mu_p-\mu_n=4.70$,
$\mu_S=\mu_p+\mu_n=0.88$ are the isovector, isoscalar moments, and
the ``extra'' factor of $1/2$ associated with the orbital angular
momentum comes from the obvious identity $\vec{L}_p=\vec{L}/2$. We
find then
\begin{eqnarray}
{e\over 2M}\mu_d&=&<\psi_d;1,1|M_3|\psi_d;1,1>={e\over
4M}\left[1+(2\mu_S-1)
<\psi_d;1,1|S_3|\psi_d;1,1>\right]\nonumber\\
&=&{e\over 4M}\left[1+(2\mu_S-1)\int d^3r(u_d^2(r)+{9\over
8}w^2(r)\left((\cos^2\theta-{1\over
3})^2-\sin^4\theta\right)\right]\nonumber\\
&=&{e\over 2M}\left[\mu_S-{3\over 2}(\mu_S-{1\over
2})\int_0^\infty drw_d^2(r) \right]\label{eq:mos}
\end{eqnarray}
In the lowest order approximation---neglecting the D-wave
component of the deuteron---we find
\begin{equation}
<1,1|M_3|1,1>\simeq \mu_S{e\over 2M}
\end{equation}
and this prediction---$\mu_d=\mu_S=0.88\mu_N$---is in good
agreement with the experimental value $\mu_d^{exp}=0.856\mu_N$.

\subsection{D-Wave Effects}

A second static observable is the quadrupole moment $Q_d$, which
is a measure of deuteron oblateness. In this case a pure S-wave
picture predicts a spherical shape so that $Q_d=0$.  Thus, in
order to generate a quadrupole moment, we must introduce a D-wave
piece of the wavefunction.  Now, just as we related the $\ell=0$
wavefunction to the np scattering in the spin triplet state, we
can relate the D-wave component to the scattering amplitude
provided we include spin.  Thus, if we write the general
scattering matrix consistent with time reversal and
parity-conservation as\cite{khr}
\begin{eqnarray}
{\cal
M}(\vec{k}',\vec{k})&=&\alpha+\beta\vec{\sigma}_p\cdot\hat{n}
\vec{\sigma}_n\cdot\hat{n}+\rho(\vec{\sigma}_p+\vec{\sigma}_n)\cdot\hat{n}
\nonumber\\
&+&(\kappa+\lambda)\vec{\sigma}_p\cdot\hat{n}_+\vec{\sigma}_n\cdot\hat{n}_-
+(\kappa-\lambda)\vec{\sigma}_p\cdot\hat{n}_-\vec{\sigma}_n\cdot\hat{n}_+
\end{eqnarray}
where
$$\hat{n}_\pm={\vec{k}\pm\vec{k}'\over |\vec{k}\pm\vec{k}'|},\qquad
\hat{n}={\vec{k}\times\vec{k}'\over |\vec{k}\times\vec{k}'|},$$ we
can represent the asymptotic scattering wavefunction via
\begin{equation}
\psi(r)\stackrel{r\rightarrow\infty}{\longrightarrow}
e^{i\vec{k}\cdot\vec{r}}+{\cal M}(-i\vec{\nabla},\vec{k})
{e^{ikr}\over r}\label{eq:gfr}
\end{equation}
A useful alternative form for ${\cal M}$ can be found via the
identity
\begin{equation}
\vec{\sigma}_p\cdot\hat{n}\vec{\sigma}_n\cdot\hat{n}=
\vec{\sigma}_p\cdot\vec{\sigma}_n-\vec{\sigma}_p\cdot\hat{n}_+
\vec{\sigma}_n\cdot\hat{n}_+-\vec{\sigma}_p\cdot\hat{n}_-\vec{\sigma}_n\cdot\hat{n}_-
\end{equation}
and, using the deuteron spin vector $\vec{S}={1\over
2}(\vec{\sigma}_p +\vec{\sigma}_n)$, it is easy to see that
\begin{equation}
{\cal M}(\vec{k}',\vec{k})=-a_t+{1\over M^2}\left[
c'\vec{S}\cdot\vec{k}\times\vec{k}'
+g_1(\vec{S}\cdot(\vec{k}+\vec{k}'))^2+g_2(\vec{S}\cdot
(\vec{k}-\vec{k}'))^2\right]
\end{equation}
where
$$c'={2\rho M^2\over k^2\sin\theta},\quad g_1={(\kappa-\beta+\lambda)M^2\over
2k^2\cos^2{\theta\over 2}},\quad
g_2={(\kappa-\beta-\lambda)M^2\over 2k^2\sin^2{\theta\over 2}} $$
Then, since in the ZRA
\begin{equation}
\vec{k}'\rightarrow -i\vec{\nabla}\delta^3(\vec{r}),\qquad\vec{k}
\rightarrow\delta^3(\vec{r})\cdot -i\vec{\nabla}
\end{equation}
we find the effective local potential
\begin{eqnarray}
U(\vec{r})&=&{4\pi\over M}\left[a_t\delta^3(\vec{r})+{c'\over M^2}
\epsilon_{ijk}S_i\nabla_j\delta^3(\vec{r})\nabla_k\right.\nonumber\\
&+&\left.{1\over
2M^2}S_{ij}\left((g_1+g_2)\{\nabla_i\nabla_j,\delta^3(\vec{r})
\}+(g_1-g_2)(\nabla_i\delta^3(\vec{r})\nabla_j+\nabla_j\delta^3(\vec{r})
\nabla_i)\right)\right]\nonumber\\
\quad
\end{eqnarray}
where
\begin{equation}
S_{ij}=S_iS_j+S_jS_i-{4\over 3}\delta_{ij}
\end{equation}
Using the Green's function representation---Eq. \ref{eq:gfr}, the
asymptotic form of the triplet scattering wavefunction becomes
\begin{equation}
\psi(r)\stackrel{r\rightarrow\infty}{\longrightarrow}e^{i\vec{k}\cdot
\vec{r}}-\left(a_t+{g_1+g_2\over
2M^2}S_{ij}\nabla_i\nabla_j\right){e^{ikr} \over r}\chi_t
\end{equation}
and, by continuing to the value $k\rightarrow i\gamma$, we can
represent the deuteron wavefunction as
\begin{equation}
\psi_d(r)\sim\sqrt{\gamma\over 2\pi}\left(1+{g_1+g_2\over 2M^2a_t}
S_{ij}\nabla_i\nabla_j\right){1\over r}e^{-\gamma r}\chi_t
\end{equation}
A little work shows that this can be written in the equivalent
form
\begin{eqnarray}
\psi_d(r)&\sim&\sqrt{\gamma\over 2\pi}\left[1+{g_1+g_2\over
2M^2a_t} {\cal O}_{pn}\left({d^2\over dr^2}-{1\over r}{d\over
dr}\right)\right]
{1\over r}e^{-\gamma r}\chi_t\nonumber\\
&=&\sqrt{\gamma\over 2\pi}\left[1+{g_1+g_2\over 2M^2a_t} {\cal
O}_{pn}\left({3\over r^2}+{3\gamma\over r}+\gamma^2\right)\right]
{1\over r}e^{-\gamma r}\chi_t
\end{eqnarray}
Here the asymptotic ratio of S- and D-wave amplitudes is an
observable and is denoted by
\begin{equation}
\eta={A_D\over A_S}={\sqrt{2}(g_1+g_2)\over 3M^2a_t^3}
\end{equation}
This quantity has been determined experimentally from elastic dp
scattering and from neutron stripping reactions to be\cite{str2}
$$\eta=0.0271\pm 0.0004,\quad i.e., \quad g_1+g_2=105\,\,{\rm fm}$$
Defining the quadrupole operator\footnote{Note that the factor of
${1\over 4}$ arises from the identity $\vec{r}_p^2={1\over
4}\vec{r}^2$}
$$Q_{ij}\equiv{e\over 4}(3r_ir_j-\delta_{ij}r^2)$$
and using
\begin{equation}
\int{d\Omega\over 4\pi}\hat{r}_i\hat{r}_j\hat{r}_k\hat{r}_\ell=
{1\over
15}(\delta_{ij}\delta_{k\ell}+\delta_{i\ell}\delta_{jk}+\delta_{ik}
\delta_{j\ell})\label{eq:qua}
\end{equation}
we note
\begin{eqnarray}
&&\int d^3r\psi_d^*(\vec{r})Q_{ij}\psi_d(\vec{r})\simeq 2e\int
d^3r\chi_t^\dagger {1\over r}u(r)Q_{ij}{(g_1+g_2)\gamma\over
2M^2}{\cal O}_{pn}\left({3\over
r^3}+{3\gamma\over r^2}+{\gamma^2\over r}\right)u(r)\chi_t\nonumber\\
&=&{3e\over 2}\cdot{1\over 15}{g_1+g_2)\gamma\over
2M^2}{\gamma\over
2\pi}\chi_t^\dagger\left(\sigma_{pi}\sigma_{nj}+\sigma_{pj}\sigma_{ni}-{2\over
3}\delta_{ij}\vec{\sigma}_p\cdot\vec{\sigma}_n\right)\chi_t\nonumber\\
&\times&4\pi\int_0^\infty
dre^{-2\gamma r}(3+3\gamma r+\gamma^2r^2)\nonumber\\
&=&{1\over 5}{e(g_1+g_2)\gamma^2\over 2M^2}\chi_t^\dagger
\left(\sigma_{pi}\sigma_{nj} +\sigma_{pj}\sigma_{ni}-{2\over
3}\delta_{ij}\vec{\sigma}_p\cdot\vec{\sigma}_n\right)\chi_t\left({3\over
2\gamma}+{3\gamma\over (2\gamma)^2}+{2\gamma^2\over
(2\gamma)^3}\right)\nonumber\\
&=&{e(g_1+g_2)\gamma\over
4M^2}\chi_t^\dagger\left(\sigma_{pi}\sigma_{nj}
+\sigma_{pj}\sigma_{ni}-{2\over
3}\delta_{ij}\vec{\sigma}_p\cdot\vec{\sigma}_n\right)\chi_t
\end{eqnarray}
Then the quadrupole moment is found to be
\begin{equation}
Q_d^{\rm th}=<\psi_d;1,1|Q_{zz}|\psi_d;1,1>={e(g_1+g_2)\over
3a_tM^2}\simeq 0.28\,\,e\,{\rm fm}^2,
\end{equation}
in good agreement with the experimental value
$$Q_d^{\rm exp}=0.286\,\,e\,{\rm fm}^2$$
From its definition, we observe that the quadrupole moment would
vanish for a spherical (purely S-wave) deuteron and that a
positive value indicates a slight elongation along the spin axis.

Note that in interpreting the meaning of the D-wave piece of the
deuteron wavefunction one often sees things described in terms of
the D-state probability
$$P_D=\int_0^\infty drw^2(r)$$
However, since
\begin{equation}
\int d^3r({3\over r^2}+{3\gamma\over r}+\gamma^2)^2\exp(-2\gamma
r)
\end{equation}
diverges while in reality the D-wave function $w_d(r)$ must vanish
as $r\rightarrow 0$, it is clear that the connection between the
asymptotic amplitude $\eta$ and the D-state probability $P_D$ must
be model dependent.  Nevertheless, the D-state piece is a small
but important component of the deuteron wavefunction.  As one
indication of this, let's return to the magnetic moment
calculation and insert the D-wave contribution. We find then
\begin{equation}
\mu_d=\mu_S-{3\over 2}(\mu_S-{1\over 2})P_D
\end{equation}
If we insert the experimental value $\mu_d=0.857$ we find
$P_D\simeq 0.04$ which can now be used in other venues.  However,
it should be kept in mind that this analysis is only approximate,
since we have neglected relativisitic corrections, meson exchange
currents, {\it etc}.

Of course, static properties represent only one type of probe of
deuteron structure.  Another is provided by the use of
electromagnetic interactions, for which a well-studied case is
photodisintegration---$\gamma d\rightarrow np$---or radiative
capture---$np\rightarrow d\gamma$---which are related by time
reversal invariance.

\section{Parity Conserving Electromagnetic Interaction: $np\leftrightarrow d\gamma$}

An important low energy probe of deuteron structure can be found
within the electromagnetic transtion $np\leftrightarrow d\gamma$.
Here the np scattering states include both spin-singlet and
-triplet components and we must include a bound state---the
deuteron.  For simplicity, we represent the latter by purely the
dominant S-wave component, which has the form
\begin{equation}
\psi_d(r)=\sqrt{\gamma\over 2\pi}{1\over r}e^{-\gamma r},\quad{\rm
or} \quad \psi_d(q)={\sqrt{8\pi\gamma}\over \gamma^2+q^2}
\end{equation}
Since we are considering an electromagnetic transition at very low
energy we can be content to include only the lowest---E1, M1, and
E2---mutipoles, which are described by the Hamiltonian\cite{khk}
\begin{equation}
H={es_0}\hat{\epsilon}_\gamma\cdot\left[\pm i{1\over
2}\vec{r}+{1\over 4M} \hat{s}_\gamma\times\left(
\mu_V(\vec{\sigma}_p-\vec{\sigma}_n)+(\mu_S-{1\over 2})
(\vec{\sigma}_p+\vec{\sigma}_s) \right) -{i\over
8}\vec{r}\vec{r}\cdot\vec{k}\right]\label{eq:ham}
\end{equation}
Here $\vec{s}_\gamma$ is the photon momentum,
$\mu_V,\mu_S=\mu_p\pm\mu_n$ are the isoscalar, isovector magnetic
moments, and we have used Siegert's theorem to convert the
conventional $\vec{p}\cdot\vec{A}/M$ interaction into the E1 form
given above.\footnote{Note that the factor of two (eight) in the
E1 (E2) component arises from the obvious identity
$\vec{r}_p\rightarrow{1\over 2}\vec{r}$\cite{khr}.} The $\pm$ in
front of the E1 operator depends upon whether the $np\rightarrow
d\gamma$ or $\gamma d\rightarrow np$ reaction is under
consideration.  The electromagnetic transition amplitude then can
be written in the form
\begin{eqnarray}
{\rm
Amp}&=&\chi_f^\dagger\left[\hat{\epsilon}_\gamma\times\hat{s}_\gamma
\cdot\left(G_{M1V}(\vec{\sigma}_p-\vec{\sigma}_n)+G_{M1S}(\vec{\sigma}_p
+\vec{\sigma}_n)\right)\right.\nonumber\\
&+&\left.G_{E1}\hat{\epsilon}_\gamma\cdot\hat{k}+G_{E2}
\left(\vec{\sigma}_p\cdot
\hat{\epsilon}_\gamma\vec{\sigma}_n\cdot\hat{s}_\gamma
+\vec{\sigma}_n\cdot\hat{\epsilon}_\gamma\vec{\sigma}_p\cdot\hat{s}_\gamma
\right)\right]\chi_i
\end{eqnarray}

The leading parity-conserving transition at near-threshold energy
is then the isovector M1 amplitude---$G_{M1V}$---which connects
the ${}^3S_1$ deuteron to the ${}^1S_0$ scattering state of the np
system. From Eq. \ref{eq:ham} we identify
\begin{equation}
G_{M1V}={es_0\mu_V\over 4M}\int
d^3r\psi_{{}^1S_0}^{(-)*}(kr)\psi_d(r)
\end{equation}
Using the asymptotic form
\begin{equation}
\psi_{{}^1S_0}^{(-)}(kr)= {e^{-i\delta_s}\over kr}(\sin
kr\cos\delta_s+\cos kr\sin\delta_s)
\end{equation}
the radial integral becomes
\begin{eqnarray}
\int d^3r\psi_{{}^1S_0}^{(-)*}(kr)\psi_d(r)&=& {4\pi
e^{i\delta_s}\over k}\int_0^\infty
(\sin kr\cos\delta_s+\cos kr\sin\delta_s)e^{-\gamma r}\nonumber\\
&=&{4\pi e^{i\delta_s}\over k(k^2+\gamma^2)}(k\cos\delta_s+ \gamma
\sin\delta_s)
\end{eqnarray}
Since by energy conservation
$$s_0={k^2+\gamma^2\over M}$$
we can use the lowest order effective range values for the
scattering phase shift to this result in the form
\begin{equation}
G_{M1V}={e\mu_V\sqrt{8\pi\gamma}e^{-i{\rm tan}^{-1}ka_s}(1-\gamma
a_s) \over 4M^2\sqrt{1+k^2a_s^2}}\label{eq:wfa}
\end{equation}
Note here that the phase of the amplitude is required by the
Fermi-Watson theorem, which follows from unitarity.

The M1 cross section is then found by squaring and mutiplying by
phase space. In the case of radiative capture this is found to be
\begin{equation}
\Gamma_{np\rightarrow d\gamma}={1\over
|\vec{v}_{rel}|}\int{d^3s\over
(2\pi)^32s_0}2\pi\delta(s_0-{\gamma^2\over M}-{k^2\over M})
\sum_{\lambda_\gamma} {1\over 4}{\rm Tr}P_tTT^\dagger
\end{equation}
Here
\begin{eqnarray}
\sum_{\lambda_\gamma}{1\over 4}{\rm Tr}P_tTT^\dagger
&=&{|G_{M1V}|^2\over 4}{\rm Tr}{1\over 4}(3+
\vec{\sigma}_1\cdot\vec{\sigma}_2)\hat{\epsilon}^*_\gamma\times\hat{s}_\gamma
\cdot(\vec{\sigma}_p-\vec{\sigma}_n)\hat{\epsilon}_\gamma\times\hat{s}_\gamma
\cdot(\vec{\sigma}_p-\vec{\sigma}_n)\nonumber\\
&=&\sum_{\lambda_\gamma}
|G_{M1V}|^2\hat{\epsilon}^*_\gamma\times\hat{s}_\gamma\cdot
\hat{\epsilon}_\gamma\times\hat{s}_\gamma=2|G_{M1V}|^2
\end{eqnarray}
yielding
\begin{equation}
\sigma_{M1}(np\rightarrow d\gamma)= {s_0\over
2\pi|\vec{v}_{rel}|}2|G_{M1V}|^2={2\pi\alpha \mu_V^2
\gamma(1-\gamma a_s)^2(k^2+\gamma^2)\over M^5(1+k^2a_s^2)}
\end{equation}
Putting in numbers we find that for an incident thermal neutrons
with relative velocity $|\vec{v}_{rel}|=2200$ m/sec. the predicted
cross section is about 300 mb which is about 10\% smaller than the
experimental value
$$\sigma_{exp}=334\pm 0.1\,\,{\rm mb}$$
The discrepancy is due to our omission of two-body effects (meson
exchange currents) as shown by Riska and Brown\cite{rib}.

\begin{figure}
\begin{center}
\epsfig{file=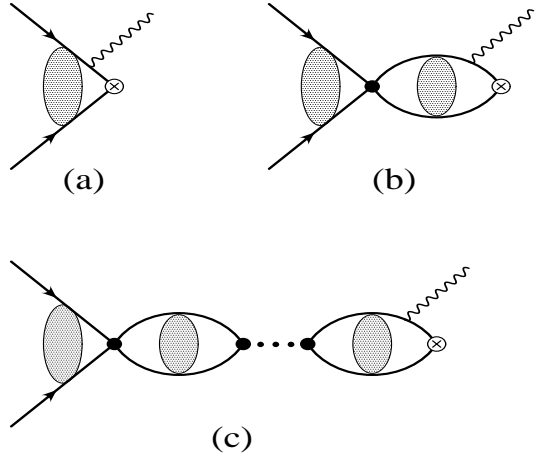,height=6cm,width=7cm} \caption{EFT
diagrams used in order to calculate the radiative capture reaction
$np\rightarrow d\gamma$.}
\end{center}
\label{fig:diag}
\end{figure}

In a corresponding EFT description of this process, we must
calculate the diagrams shown in Figure 1.  There is a subtlety
here which should be noted. Strictly speaking, as shown by Kaplan,
Savage, and Wise\cite{ksw} the symbol $\otimes$ in these diagrams
should be interpreted as creation or annihilation of the deuteron
with wavefunction renormalization
\begin{equation}
\sqrt{Z}=\left({d\Sigma(E)\over dE}\right)_{E=-B}^{-{1\over 2}}
\end{equation}
followed by propagation via
$${1\over {k^2\over M}+{\gamma^2\over M}}$$
However, since in lowest order we have
\begin{equation}
\left({d\Sigma(E)\over dE}\right)_{E=-B}^{-{1\over 2}}=
{\sqrt{8\pi\gamma}\over M}
\end{equation}
we find for the product
\begin{equation}
\sqrt{Z}\cdot{1\over {k^2\over M}+{\gamma^2\over
M}}={\sqrt{8\pi\gamma}\over k^2+\gamma^2}
\end{equation}
which is the deuteron wavefunction in momentum space.  Thus in our
discussion below we shall use this substitution rather that
writing the wavefunction normalization times propagator product.
From Figure 4a then we find
\begin{equation}
G_{M1V}^a={es_0\mu_V\over 4M} \int{d^3q\over
(2\pi)^3}\psi_{\vec{k}}^{(0)*}(\vec{q}) \psi_d(\vec{q})
\end{equation}
Since
$\psi_{\vec{k}}^{(0)*}(\vec{q})=(2\pi)^3\delta^3(\vec{k}-\vec{q})$
we have
\begin{equation}
G_{M1V}^a={es_0\mu_V\over 4M}{\sqrt{8\pi\gamma}\over \gamma^2+k^2}
\end{equation}
On the other hand from Figures 4b+c we find
\begin{equation}
G_{M1V}^{b+c}={es_0\mu_V\over 4M}\left({C_{0s}\over
1-C_{0s}G_0(k)}\right)^* \int{d^3q\over
(2\pi)^3}G_0(\vec{r}=0,\vec{q})\psi_d(q)
\end{equation}
Since
\begin{equation}
G_0(\vec{r}=0,\vec{q})={1\over {k^2\over M}-{q^2\over
M}+i\epsilon}
\end{equation}
this becomes
\begin{eqnarray}
G_{M1V}^{b+c}&=&{es_0\mu_V\over 4M}\left({C_{0s}\over
1-C_{0s}G_0(k)}\right)^* \int{d^3q\over (2\pi)^3}
{\sqrt{8\pi\gamma}\over (q^2+\gamma^2)
({k^2\over M}-{q^2\over M}+i\epsilon)}\nonumber\\
&=&{es_0\mu_V\over 4M}\left({C_{0s}\over 1-C_{0s}G_0(k)}\right)^*
{\sqrt{8\pi\gamma}\over
\gamma^2+k^2}(G_0(k)-G_0(i\gamma))\nonumber\\
&=&{es_0\mu_V\over 4M}\left({C_{0s}\over 1-C_{0s}G_0(k)}\right)^*{
\sqrt{8\pi\gamma}\over \gamma^2+k^2} {M\over 4\pi}(-ik-\gamma)
\end{eqnarray}
Adding the two contributions we have
\begin{eqnarray}
G_{M1V}&=&{es_0\mu_V\over 4M}{\sqrt{8\pi\gamma}\over
\gamma^2+k^2}\left[
1-\left({-\gamma-ik\over {-1\over a_s}-ik}\right)\right]\nonumber\\
&=&{es_0\mu_V\over 4M}{\sqrt{8\pi\gamma}\over
\gamma^2+k^2}\left({1-\gamma a_s\over 1+ika_s} \right)
\end{eqnarray}
which agrees completely with Eq. \ref{eq:wfa} obtained via
conventional coordinate space procedures.  Of course, we still
have a $\sim 10\%$ discrepancy with the experimental cross
section, which is handled by inclusion of a four-nucleon M1
counterterm connecting ${}^3S_1$ and ${}^1S_0$ states---
\begin{equation}
{\cal
L}_2^{EM}=eL_1^{M1V}(N^T\vec{P}\cdot\vec{B}N)^\dagger(N^TP_3N)
\end{equation}
where here the $P_i$ represent relvant projection
operators\cite{rup}.

As the energy increases above threshold, it is necessary to
include the corresponding P-conserving E1 multipole---
\begin{equation}
{\rm
Amp}=G_{E1}\hat{\epsilon}_\gamma\cdot\vec{k}\chi_t^\dagger\chi_t
\end{equation}
In this case the matrix element involves the np $^3$P-wave final
state and, neglecting final state interactions in this channel,
the matrix element is given by
\begin{equation}
G_{E1}\vec{k}={es_0\over 2}\int
d^3r\psi_{\vec{k}}^{(0)*}(r)\vec{r}\psi_d(r)
\end{equation}
The radial integral can be found via
\begin{eqnarray}
-i\vec{k}\cdot\int d^3r
e^{-i\vec{k}\cdot\vec{r}}\vec{r}\psi_d(r)&=& {d\over
d\lambda}_{|_{\lambda=1}}\sqrt{\gamma\over 2\pi}\int d^3r
e^{-i\vec{k}\cdot\vec{r}\lambda}{1\over r}e^{-\gamma r}\nonumber\\
&=&{d\over d\lambda}_{|_{\lambda=1}}{\sqrt{8\pi\gamma}\over
\gamma^2+k^2\lambda^2}={-2k^2\sqrt{8\pi\gamma}\over
(k^2+\gamma^2)^2}
\end{eqnarray}
Equivalently using EFT methods we have
\begin{equation}
\int {d^3q\over (2\pi)^3}\psi^{(0)*}_{\vec{k}}(\vec{q})
\vec{\nabla}_{\vec{q}}\psi_d(q) ={-2i\sqrt{8\pi\gamma}\vec{k}\over
(k^2+\gamma^2)^2}
\end{equation}
In either case
\begin{equation}
G_{E1}={-ies_0\sqrt{8\pi\gamma}\over (k^2+\gamma^2)^2}
\end{equation}
and the corresponding cross section is
\begin{eqnarray}
\sigma_{E1}(np\rightarrow d\gamma)&=&{1\over 4}{3s_0\over
2\pi|\vec{v}_{rel}|} \int {d\Omega_{\hat{s}}\over 4\pi}
(\vec{k}\cdot\vec{k}-(\vec{k}\cdot\hat{s})^2){e^28\pi\gamma
s_0^2\over
(k^2+\gamma^2)^4}\nonumber\\
&=&{8\pi\alpha k^2\gamma\over |\vec{v}_{rel}|M^3(k^2+\gamma^2)}
\end{eqnarray}

It is important to note that we can easily find the corresponding
photodisintegration cross sections by multiplying by the
appropriate phase space.  For unpolarized photons we have
\begin{eqnarray}
\sigma(\gamma d\rightarrow np)&=&{1\over 3\cdot 2}{1\over
2s_0}\int {d^3k\over
(2\pi)^3}\sum_{\lambda_\gamma}2\pi\delta(s_0-{\gamma^2\over M}
-{k^2\over M})|{\rm Amp}|^2\nonumber\\
&=&{Mk\over 24\pi
s_0}\sum_{\lambda_\gamma}\int{d\Omega_{\hat{k}}\over 4\pi} |{\rm
Amp}|^2
\end{eqnarray}
Using the results obtained above for radiative capture---
\begin{equation}
\sum_{\lambda_\gamma}|{\rm
Amp}|^2=8|G_{M1V}|^2+3(\vec{k}\cdot\vec{k}
-(\vec{k}\cdot\hat{s})^2)|G_{E1}|^2
\end{equation}
we find the photodisintegration cross sections to be
\begin{equation}
\sigma_{M1}(\gamma d\rightarrow np)= {2\pi\alpha\over
3M^2}{(1-a_s\gamma)^2\mu_V^2k\gamma\over
(k^2+\gamma^2)(1+k^2a_0^2)},\quad \sigma_{E1}(\gamma d\rightarrow
np) ={8\pi\alpha\gamma k^3\over 3(k^2+\gamma^2)^3}
\end{equation}

Although the leading electromagnetic physics is controlled, as we
have seen, by the isovector M1 and E1 amplitudes, there exist
small but measurable isoscalar M1 and E2
transitions\cite{rus},\cite{sea},\cite{seo}.  In the former case
the transition is between the S-wave (D-wave) deuteron ground
state and into the ${}^3S_1 ({}^3D)$ scattering state.  The
amplitude $G_{M1S}$ is small because of the smallness of
$\mu_S-{1\over 2}$ and because of the orthogonality restriction.
In the case of $G_{E2}$, the result is suppressed by the
requirement for transfer of two units of angular momentum, so that
the transition must be between S- and D-wave components of the
wavefunction.

We first evaluate the isoscalar M1 amplitude, which from Eq.
\ref{eq:ham} is given by ({\it cf.} Eq. \ref{eq:mos})
\begin{eqnarray}
G_{M1S}&=&{e\over 2M}(\mu_S-{1\over
2})<\psi_d;1,1|S_3|\psi_d,1,1>\nonumber\\
&=&{e\over 2M}(\mu_S-{1\over
2})\int_0^\infty dr(u_d(r)u_t(r)-{1\over 2}w_d(r)w_t(r))\nonumber\\
&=&-{e\over 2M}(\mu_S-{1\over 2}){3\over 2}\int_0^\infty dr
w_t(r)w_d(r)
\end{eqnarray}
where the last form was fouind using the orthogonality condition.
In order to estimate the latter, we follow Danilov in assuming
that, since the radial integral is short-distance dominated, the
D-wave deuteron and scattering pieces are related by a simple
constant\cite{dan}, which, using orthogonality, must be given by
\begin{equation}
w_t(r)\simeq -a_t\sqrt{2\pi\over \gamma}w_d(r)
\end{equation}
The matrix element then becomes
\begin{equation}
G_{M1S}\simeq {e\over 2M}(\mu_S-{1\over 2}){3\over 2}P_Da_t
\end{equation}

Likewise, using Eq. \ref{eq:qua} we note that
\begin{eqnarray}
&&\int d^3r \left({1\over r}u_t(r)+{3\over \sqrt{8}}{\cal
O}_{pn}{1\over r} w_t(r)\right)\vec{r}
\cdot\hat{\epsilon}_\gamma\vec{r}\cdot\hat{s}_\gamma\psi_d(\vec{r})\nonumber\\
&=&{3\over \sqrt{8}}{1\over 15}\int_0^\infty
drr^2(u_t(r)w_d(r)+w_t(r)u_d(r))
\left[\vec{\sigma}_p\cdot\hat{\epsilon}_\gamma\vec{\sigma}_n
\cdot\hat{s}_\gamma+\vec{\sigma}_n\cdot\hat{\epsilon}_\gamma\vec{\sigma}_p
\cdot\hat{s}_\gamma\right]\nonumber\\
\quad
\end{eqnarray}
so that the corresponding E2 coupling is found to be
\begin{equation}
G_{E2}={es_0\over 80\sqrt{2}}\int_0^\infty
drr^2(u_t(r)w_d(r)+w_t(r)u_d(r))=-{es_0\over
80\sqrt{2}}{g_1+g_2\over 2M^2}
\end{equation}

In order to detect these small components we can use the circular
polarization induced in the final state photon by an initially
polarized neturon, which is found to be
\begin{equation}
P_\gamma=-2\left({G_{M1S}-G_{E2}\over G_{M1V}}\right)
\end{equation}
Putting in numbers we find
\begin{eqnarray}
P_\gamma&=&P_\gamma(M1)+P_\gamma(E2)\nonumber\\
&=&{-\gamma a_t\over \mu_V(1-\gamma a_s)}\left(\mu_S-\mu_d+{2\over
15}{\gamma^2(g_1+g_2)\over
a_tM^2}\right)\nonumber\\
&\simeq&-1.17\times 10^{-3}-0.24\times 10^{-3}\nonumber\\
&=&-1.41\times 10^{-3}
\end{eqnarray}
which is in reasonable agreement with the experimental
value\cite{exp}
\begin{equation}
P_\gamma^{exp}=(-1.5\pm 0.3)\times 10^{-3}
\end{equation}

Having familiarized ourselves with the analytic techniques which
are needed, we now move to our main subject, which is hadronic
parity violation in the NN system.

\section{Parity-Violating NN Scattering}

For simplicity we begin again with a system of two nucleons. Then
the NN scattering-matrix can be written at low energies in the
phenomenological form\cite{dan}
\begin{equation}
{\cal M}(\vec{k}',\vec{k})=m_t(k)P_1+m_s(k)P_0\label{eq:tmx}
\end{equation}
where
$$P_1={1\over 4}(3+\vec{\sigma}_1\cdot\vec{\sigma}_2),\qquad
P_0={1\over 4}(1-\vec{\sigma}_1\cdot\vec{\sigma}_2)$$ are
spin-triplet, -singlet spin projection operators and
\begin{equation}
m_t(k)={-a_t\over 1+ika_t},\qquad m_s(k) ={-a_s\over 1+ika_s}
\end{equation}
are the S-wave partial wave amplitudes in the lowest order
effective range approximation, keeping only the scattering lengths
$a_t,a_s$.  Here the scattering cross section is found via
\begin{equation}
{d\sigma\over d\Omega}={\rm Tr}{\cal M}^\dagger{\cal M}
\end{equation}
so that at the lowest energy we have the familiar form
\begin{equation}
{d\sigma_{s,t}\over d\Omega}={|a_{s,t}|^2\over 1+k^2a_{s,t}^2}
\end{equation}
The corresponding scattering wavefunctions are then given by
\begin{eqnarray}
\psi^{(+)}_{\vec{k}}(\vec{r})&=&\left[e^{i\vec{k}\cdot\vec{r}}-{M\over
4\pi} \int d^3r'{e^{ik|\vec{r}-\vec{r}'|}\over
|\vec{r}-\vec{r}'|}U(\vec{r}')
\psi^{(+)}_{\vec{k}}(\vec{r})\right]\chi\nonumber\\
&&\stackrel{r\rightarrow\infty}{\longrightarrow}
\left[e^{i\vec{k}\cdot\vec{r}}+{\cal M}(-i\vec{\nabla},
\vec{k}){e^{ikr}\over r}\right]\chi\label{eq:wfc}
\end{eqnarray}
where $\chi$ is the spin function. In Born approximation we can
represent the wavefunction in terms of an effective delta function
potential
\begin{equation}
U_{t,s}(\vec{r})={4\pi\over M}(a_tP_1+a_sP_0)\delta^3(\vec{r})
\end{equation}
as can be confirmed by substitution into Eq. \ref{eq:wfc}.

\subsection{Including the PV Interaction}

Following Danilov,\cite{dan}, we can introduce parity mixing into
this simple representation by generalizing the scattering
amplitude to include P-violating structures. Up to laboratory
energies of 50 MeV or so, we can omit all but S- and P-wave
mixing, in which case there exist only five independent such
amplitudes:
\begin{itemize}
\item[i)] $d_t(k)$ representing ${}^{3}S_1--{}^{1}P_1$ mixing;
\item[ii)] $d_s^{0,1,2}(k)$ representing ${}^{1}S_0--{}^{3}P_0$ mixing
with $\Delta I=0,1,2$ respectively;
\item[iii)] $c_t(k)$ representing ${}^{3}S_1--{}^{3}P_1$ mixing.
\end{itemize}
After a little thought, it becomes clear then that the low energy
scattering-matrix in the presence of parity violation can be
written as
\begin{eqnarray}
{\cal M}(\vec{k}',\vec{k})&=&\left[m_s(k)P_0+c_t(k)(\vec{\sigma}_1
+\vec{\sigma}_2)\cdot
(\vec{k}'+\vec{k}){1\over 2}(\tau_1-\tau_2)_z\right.\nonumber\\
&+&\left.(\vec{\sigma}_1-\vec{\sigma}_2)\cdot
(\vec{k}'+\vec{k})\left(P_0d_s^0(k)+{1\over
2}(\tau_1+\tau_2)_zd_s^1(k)
+{3\tau_{1z}\tau_{2z}-\vec{\tau}_1\cdot\vec{\tau}_2\over
2\sqrt{6}}d_s^2(k)
\right)\right]\nonumber\\
&+&\left[m_t(k)+d_t(k)(\vec{\sigma}_1-\vec{\sigma}_2)\cdot
(\vec{k}'+\vec{k})\right]P_1\label{eq:tpx}
\end{eqnarray}
Note that since under spatial
inversion---$\vec{\sigma}\rightarrow\vec{\sigma},
\vec{k},\vec{k}'\rightarrow-\vec{k},-\vec{k}'$---each of the new
pieces is P-odd, and since under time reversal---$\vec{\sigma}
\rightarrow-\vec{\sigma},
\vec{k},\vec{k}'\rightarrow-\vec{k}',-\vec{k}$ the terms are each
T-even. At very low energies the coefficients in the T-matrix
become real and we define\cite{dan}
\begin{equation}
\lim_{k\rightarrow 0}m_{s,t}(k)=a_{s,t},\quad \lim_{k\rightarrow
0}c_t(k), d_s(k),d_t(k)=\rho_ta_t,\lambda_s^ia_s,\lambda_ta_t
\end{equation}
(The reason for factoring out the S-wave scattering length will be
described presently.)  The five real numbers
$\rho_t,\lambda_s^i,\lambda_t$ then completely characterize the
low energy parity-violating interaction and can in principle be
determined experimentally, as we shall discuss
below.\footnote{Note that there exists no singlet analog to the
spin-triplet constant $c_t$ since the combination
$\vec{\sigma}_1+\vec{\sigma}_2$ is proportional to the total spin
operator and vanishes when operating on a spin singlet state.}
Alternatively, we can write things in terms of the equivalent
notation
\begin{eqnarray}
\lambda_s^{pp}&=&\lambda_s^0+\lambda_s^1+{1\over
\sqrt{6}}\lambda_s^2\nonumber\\
\lambda_s^{np}&=&\lambda_s^0-{2\over \sqrt{6}}\lambda_s^2\nonumber\\
\lambda_s^{nn}&=&\lambda_s^0-\lambda_s^1+{1\over
\sqrt{6}}\lambda_s^2
\end{eqnarray}

We can also represent this interaction in terms of a simple
effective NN potential.  Integrating by parts, we have
\begin{equation}
\int d^3r'{e^{ik|\vec{r}-\vec{r}'|}\over |\vec{r}-\vec{r}'|}
\{-i\vec{\nabla},\delta^3(\vec{r}')\}
e^{i\vec{k}\cdot\vec{r}'}=(-i\vec{\nabla}+\vec{k}){e^{ikr}\over r}
\end{equation}
which represents the parity-violating admixture to the the
scattering wavefunction in terms of an S-wave admixture to the
scattering P-wave state---$\sim
\vec{\sigma}\cdot\vec{k}{e^{ikr}\over r}$ plus a P-wave admixture
the scattering S-state---$\sim
-i\vec{\sigma}\cdot\vec{\nabla}{e^{ikr}\over r}$.  We see then
that the scattering wave function can be described via
\begin{eqnarray}
U(\vec{r})&=&{4\pi\over M}\left[\left(a_t\delta^3(\vec{r})+
\lambda_ta_t(\vec{\sigma}_1-\vec{\sigma}_2)\cdot
\{-i\vec{\nabla},\delta^3(\vec{r})\}\right)P_1\right.\nonumber\\
&+&\left.a_s\delta^3(\vec{r})P_0+\rho_ta_t(\vec{\sigma}_1+
\vec{\sigma}_2)\cdot\{-i\vec{\nabla},\delta^3(\vec{r})\}{1\over 2}
(\tau_1-\tau_2)_z\right.\nonumber\\
&+&\left.(\vec{\sigma}_1-\vec{\sigma}_2)\cdot
\{-i\vec{\nabla},\delta^3(\vec{r})\}a_s\left(P_0\lambda_s^0+{1\over
2} (\tau_1+\tau_2)_z\lambda_s^1
+{3\tau_{1z}\tau_{2z}-\vec{\tau}_1\cdot\vec{\tau}_2\over
2\sqrt{6}}\lambda_s^2
\right)\right]\nonumber\\
\qquad\label{eq:eff}
\end{eqnarray}
However, before application of this effective potential we must
worry about the stricture of unitarity, which requires that
\begin{equation}
2{\rm Im}T=T^\dagger T
\end{equation}
In the case of the S-wave partial wave amplitude $m_t(k)$ this
condition reads
\begin{equation}
{\rm Im}\,m_{t}(k)=k|m_{t}(k)|^2
\end{equation}
and requires the form
\begin{equation}
m_{t}(k)={1\over k}e^{i\delta_{t}(k)}\sin\delta_{t}(k)
\end{equation}
Since at zero energy we have
\begin{equation}
\lim_{k\rightarrow 0}m_{t}(k)=-a_{t}
\end{equation}
It is clear that unitarity can be enforced by modifying this
lowest order result via
\begin{equation}
m_{t}(k)={-a_{t}\over 1+ika_{t}}\label{eq:noo}
\end{equation}
which is the lowest order effective range result.  Equivalently,
this can easily be derived in an effective field theory (EFT)
formalism. In this case the lowest order contact interaction
\begin{equation}
T_{0t}=C_{0t}(\mu)
\end{equation}
becomes, when summed to all orders in the scattering series,
\begin{equation}
T_t(k)={C_{0t}(\mu)\over 1-C_{0t}(\mu)G_0(k)}=-{M\over 4\pi}
{1\over -{4\pi\over MC_{0t}(\mu)}-\mu-ik}\label{eq:not}
\end{equation}
Identifying the scattering length via
\begin{equation}
-{1\over a_t}=-{4\pi\over MC_{0t}(\mu)}-\mu
\end{equation}
and noting the relation $m_{t}(k)=-{M\over 4\pi}T_t(k)$ connecting
the scattering and transition matrices, we see that Eqs.
\ref{eq:noo} and \ref{eq:not} are identical.

So far, so good.  However, things become more interesting in the
case of the parity-violating transitions.  In this case the
requirement of unitarity reads, {\it e.g.}, for the case of
scattering in the ${}^3S_1$ channel
\begin{equation}
{\rm Im}\,d_t(k)=k(m_t^*(k)d_t(k)+d_t^*(k)m_p(k))\label{eq:uni}
\end{equation}
where $m_p(k)$ is the ${}^1P_1$ analog of the $m_t(k)$.  Eq.
\ref{eq:uni} is satisfied by the solution
\begin{equation}
d_t(k)=|d_t(k)|e^{i(\delta_{{}^3S_1}(k)+\delta_{{}^1P_1}(k))}
\end{equation}
{\it i.e.}, the phase of the amplitude should be the sum of the
strong interaction phases in the incoming and outgoing
channels\cite{mil}. At very low energy we can neglect P-wave
scattering, and can write
\begin{equation}
c_t(k)\simeq\rho_tm_t(k),\quad
d_s^i(k)\simeq\lambda_s^im_s(k),\quad d_t(k) \simeq\lambda_tm_t(k)
\end{equation}
This result is also easily seen in the language of EFT, wherein
the full transition matrix must include the weak amplitude to
lowest order accompanied by rescattering in both incoming and
outgoing channels to all orders in the strong interaction.  If we
represent the lowest order weak contact interaction as
\begin{equation}
T_{0tp}(k)=D_{0tp}(\mu)(\vec{\sigma}_1-\vec{\sigma}_2)\cdot(\vec{k}+\vec{k}')
\end{equation}
then the full amplitude is given by
\begin{equation}
T_{tp}(k)={D_{0tp}(\mu)\over
(1-C_{0t}(\mu)G_0(k))(1-C_{0p}(\mu)G_1(k))}
(\vec{\sigma}_1-\vec{\sigma}_2)\cdot(\vec{k}+\vec{k}')
\end{equation}
where we have introduced a lowest order contact term $C_{0p}$
which describes the ${}^1P_1$-wave nn interaction.  Since the
phase of the combination $1-C_0(\mu)G_0(k)$ is simply the negative
of the strong interaction phase the unitarity stricture is clear,
and we can define the physical transition amplitude $A_{tp}$ via
\begin{equation}
{D_{0tp}(\mu)\over (1-C_{0t}(\mu)G_0(k))(1-C_{0p}(\mu)G_1(k))}
\equiv {A_{tp}\over (1+ika_t)(1+ik^3a_p)}
\end{equation}
Making the identification $\lambda_t= -{M\over 4\pi}A_{tp}$ and
noting that $${1\over 1+ika_t}=\cos\delta_t(k)e^{i\delta_t(k)}$$
then $\lambda_t$ is seen to be identical to the R-matrix element
defined by Miller and Driscoll\cite{mil}.

Now that we have developed a fully unitary transition amplitude we
can calculate observables.  For simplicity we begin with nn
scattering.  In this case the Pauli principle demands that the
initial state must be purely ${}^1S_0$ at low energy. One can
imagine longitudinally polarizing one of the neutrons and
measuring the total scattering cross section on an unpolarized
target. Since $\vec{\sigma}\cdot\vec{k}$ is odd under parity, the
cross section can depend on the helicity only if parity is
violated.  Using trace techniques the helicity correlated cross
section can easily be found. Since the initial state must be in a
spin singlet we have
\begin{eqnarray}
\sigma_\pm&=&\int d\Omega{1\over 2}{\rm Tr}{\cal
M}(\vec{k}',\vec{k}) {1\over
2}(1+\vec{\sigma}_2\cdot\hat{k}){1\over 4}(1-\vec{\sigma}_1\cdot
\vec{\sigma}_2){\cal M}^\dagger(\vec{k}',
\vec{k})\nonumber\\
&=&|m_s(k)|^2\pm 4k{\rm Re}\,m_s^*(k)d_s^{nn}(k)+{\cal O}(d_s^2)
\end{eqnarray}
Defining the asymmetry via the sum and difference of such helicity
cross sections and neglecting the tiny P-wave scattering, we have
then
\begin{equation}
A={\sigma_+-\sigma_-\over \sigma_++\sigma_-}={8k{\rm
Re}\,m_s^*(k)d_s^{nn}(k) \over
2|m_s(k)|^2}=4k\lambda_s^{nn}\label{eq:fre}
\end{equation}
Thus the helicity correlated nn-scattering asymmetry provides a
direct measure of the parity-violating parameter $\lambda_s^{nn}$.
Note that in the theoretical evaluation of the asymmetry, since
the total cross section is involved some investigators opt to
utilize the optical theorem via\cite{bhk},\cite{oka}
\begin{equation}
A={4k{\rm Im}\,d_s^{nn}(k)\over {\rm Im}\,m_s(k)}
\end{equation}
which, using our unitarized forms, is completely equivalent to Eq.
\ref{eq:fre}.

Of course, nn-scattering is purely a gedanken experiment and we
have discussed it only as a warmup to the real problem---pp
scattering, which introduces the complications associated with the
Coulomb interaction.  In spite of this complication, the
calculation proceeds quite in parallel to the discussion above
with obvious modifications.  Specifically, as shown in \cite{hol}
the unitarized the scattering amplitude now has the form
\begin{equation}
m_s(k)=-{M\over 4\pi}{C_{0s}C_\eta^2(\eta_+(k))\exp2i\sigma_0\over
1-C_{0s}G_C(k)}
\end{equation}
where $\eta_+(k)=M\alpha/2k$ and
\begin{equation}
C^2(x)={2\pi x\over e^{2\pi x}-1}
\end{equation}
is the usual Sommerfeld factor and $\sigma_0= {\rm
arg}\Gamma(\ell+1+i\eta(k))$ is the Coulomb phase shift.  Of
course, the free Green's function $G_0(k)$ has also been replaced
by its Coulomb analog
\begin{equation}
G_C(k)=\int{d^3s\over (2\pi)^3}{C^2(\eta_+(k))\over {k^2\over
M}-{s^2\over M}+i\epsilon}
\end{equation}
Remarkably this integral can be performed analytically and the
result is
\begin{equation}
G_C(k)=-{M\over
4\pi}\left[\mu+M\alpha\left(H(i\eta_+(k))-\log{\mu\over \pi
M\alpha}-\zeta\right)\right]
\end{equation}
Here $\zeta$ is defined in terms of the Euler constant $\gamma_E$
via $\zeta=2\pi-\gamma_E$ and
\begin{equation}
H(x)=\psi(x)+{1\over 2x}-\log x
\end{equation}
The resultant scattering amplitude has the form
\begin{eqnarray}
m_s(k)&=&{C_\eta^2(\eta_+(k))e^{2i\sigma_0}\over -{4\pi\over
MC_{0s}} -\mu-M\alpha\left[H(i\eta_+(k))-\log{\mu\over \pi
M\alpha}
-\zeta\right]}\nonumber\\
&=&{C_\eta^2(\eta_+(k))e^{2i\sigma_0}\over -{1\over a_{0s}}
-M\alpha\left[h(\eta_+(k))-\log{\mu\over \pi M\alpha}
-\zeta\right]-ikC_\eta^2(\eta_+(k))}
\end{eqnarray}
where we have defined
\begin{equation}
-{1\over a_{0s}}=-{4\pi\over MC_{0s}}-\mu,\quad{\rm and}\quad
h(\eta_+(k)) ={\rm Re}H(i\eta_+(k))
\end{equation}
The experimental scattering length $a_{Cs}$ in the presence of the
Coulomb interaction is defined via
\begin{equation}
-{1\over a_{Cs}}=-{1\over a_{0s}}+M\alpha\left(\log{\mu\over \pi
M\alpha} -\zeta\right)\label{eq:sca}
\end{equation}
in which case the scattering amplitude takes its traditional form
\begin{equation}
m_s(k)={C_\eta^2(\eta_+(k))e^{2i\sigma_0}\over -{1\over a_{Cs}}
-M\alpha H(i\eta_+(k))}
\end{equation}
Of course, this means that the Coulomb-corrected scattering length
is different from its non-Coulomb analog, and comparison of the
experimental pp scattering length---$a_{pp}=-7.82$ fm---with its
nn analog---$a_{nn}= -18.8$ fm---is roughly consistent with Eq.
\ref{eq:sca} if a reasonable cutoff, say $\mu\sim 1 GeV$ is
chosen.  Having unitarized the strong scattering amplitude, we can
proceed similarly for its parity-violating analog.  Again summing
the rescattering bubbles and neglecting the small p-wave
scattering, we find for the unitarized weak amplitude
\begin{equation}
T_{0SP}={D_{0sp}(\mu)C_\eta^2(\eta_+(k))e^{i(\sigma_0+\sigma_1)}
\over 1-C_{0s}(\mu)G_C(k)} \equiv
{A_{Csp}C_\eta^2(\eta_+(k))e^{i(\sigma_0+\sigma_1)}\over -{1\over
a_{Cs}} -M\alpha H(i\eta_+(k))}
\end{equation}
Here again, the Driscoll-Miller procedure identifies
$A_{Csp}=\lambda_s^{pp}$ via the R-matrix. Having obtained fully
unitarized forms, we can then proceed to evaluate the helicity
correlated cross sections, finding as before
\begin{equation}
A_h={\sigma_+-\sigma_-\over \sigma_++\sigma_-}={8k{\rm
Re}\,m_s^*(k)d_s^{pp}(k) \over 2|m_s(k)|^2}\simeq 4k\lambda_s^{pp}
\end{equation}
Note here that the superscript $pp$ has been added, in order
account for the feature that in the presence of Coulomb
interactions the parity mixing parameter $\lambda_s$ which is
appropriate for neutral scattering is modified, in much the same
way as the scattering length in the $pp$ channel is modified ({\it
cf.} Eq. \ref{eq:sca}). On the experimental side such asymmetries
have been measured both at low energy (13.6 and 45 MeV) as well as
at higher energy (221 and 800 MeV) but it is only the low energy
results\footnote{Note that the 13.6 MeV Bonn measurement is fully
consistent with the earlier but less precise number
\begin{equation}
A_h=-(1.7\pm 0.8)\times 10^{-7}\cite{lan}
\end{equation}
determined at LANL.}
\begin{eqnarray}
A_h(13.6\,\,{\rm MeV})&=&-(0.93\pm0.20\pm0.05)\times
10^{-7}\cite{bon}\nonumber\\
A_h(45\,\,{\rm MeV})&=&-(1.57\pm 0.23)\times 10^{-7}\cite{psi}
\end{eqnarray}
which are appropriate for our analysis. Note that one consistency
check on these results is that if the simple discussion given
above is correct the two numbers should be approximately related
by the kinematic factor\footnote{There is an additional
$k$-dependence arising from $\lambda_s$ but this is small.}
\begin{equation}
A_h(45\,\,{\rm MeV})/A_h(13.6\,\,{\rm MeV})\simeq k_1/k_2=1.8
\end{equation}
and the quoted numbers are quite consistent with this requirement.
We can then extract the experimental number for the singlet mixing
parameter as
\begin{equation}
\lambda_s^{pp}={A_h\over 4k}=-(4.0\pm 0.8)\times 10^{-8}\,\,{\rm
fm} \label{eq:ppe}
\end{equation}
In principle one could extract the triplet parameters by a careful
nd scattering measurement.  However, extraction of the neeeded np
amplitude involves a detailed theoretical analysis which has yet
not been performed. Thus instead we discuss the case of
electromagnetic interactions and consider $np\leftrightarrow
d\gamma$.

\section{Parity Violating Electromagnetic Interaction: $np\leftrightarrow d\gamma$}

A second important low energy probe of hadronic parity violation
can be found within the electromagnetic transtion
$np\leftrightarrow d\gamma$. Here the np scattering states include
both spin-singlet and -triplet components and we must include a
bound state---the deuteron.  Analysis of the corresponding
parity-conserving situation has been given previously, so we
concentrate here on the parity violating situation. In this case,
the mixing of the scattering states has already been given in Eq.
\ref{eq:tpx} while for the deuteron the result can be found from
demanding orthogonality with the ${}^3S_1$ scattering state---
\begin{equation}
\psi_d(r)=\left(1+\rho_t(\vec{\sigma}_p+\vec{\sigma}_n)
\cdot-i\vec{\nabla}
+\lambda_t(\vec{\sigma}_p-\vec{\sigma}_n)\cdot-i\vec{\nabla}\right)
\sqrt{\gamma\over 2\pi}{1\over r}e^{-\gamma r}\label{eq:deu2}
\end{equation}
Having found $\lambda_s$ via the pp scattering asymmetry, we now
need to focus on the determination of the parity-violating triplet
parameters $\rho_t,\lambda_t^i$.  In order to do so, we must
evaluate new matrix elements. There are in general two types of PV
E1 matrix elements, which we can write as
\begin{equation}
{\rm
Amp}=\left(H_{E1}\chi_s^\dagger(\vec{\sigma}_p-\vec{\sigma}_n)\chi_t
+S_{E1}\chi_t^\dagger(\vec{\sigma}_p+\vec{\sigma}_n)\chi_t\right)
\cdot\hat{\epsilon}_\gamma
\end{equation}
and there exist two separate contributions to each of these
amplitudes, depending upon whether the parity mixing occurs in the
inital or final state.  We begin with the matrix element which
connects the ${}^1P_1$ admixture of the deuteron to the ${}^1S_0$
scattering state.
\begin{eqnarray}
H_{E1}({}^1P-{}^1S)&=&{es_0\lambda_t\over 2} {1\over 3}\int d^3r
\psi_{{}^1S}^{(-)*}(r)
\vec{r}\cdot\vec{\nabla}\psi_d(r)\nonumber\\
&=&{es_04\pi\lambda_t\over 6}e^{i\delta_s}\int_0^\infty
drr^2{1\over kr} (\sin kr\cos \delta_s+\cos kr\sin
\delta_s)\vec{r}\cdot\vec{\nabla}
\sqrt{\gamma\over 2\pi}{1\over r}e^{-\gamma r}\nonumber\\
&=&{es_0\sqrt{8\pi\gamma}\lambda_t\over 6}
{e^{i\delta_s}\over k}\nonumber\\
&\times&(1-\gamma{d\over d\gamma}) \int_0^\infty dr (\sin kr
\cos\delta_s+\cos kr\sin\delta_s)
e^{-\gamma r}\nonumber\\
&=&{es_0\lambda_t\sqrt{8\pi\gamma}e^{-i{\rm tan}^{-1}ka_s}
\over 6\sqrt{1+k^2a_s^2}}\nonumber\\
&\times& \left[{k^2+3\gamma^2\over (k^2+\gamma^2)^2}-\gamma
a_s{2\gamma^2\over (k^2+\gamma^2)^2}\right]
\end{eqnarray}
Equivalently we can use EFT methods using the diagrams of Figure
4. We have from Figure 4a
\begin{eqnarray}
H_{E1}^{a}({}^1P-{}^1S)&=&{es_0\lambda_t\over 2}{1\over 3}\int
{d^3q\over
(2\pi)^3}\psi_{\vec{k}}^{(0)*}(\vec{q})\vec{\nabla}_{\vec{q}}
\cdot\left(
\vec{q}\psi_d(q)\right)\nonumber\\
&=&{es_0\lambda_t\sqrt{8\pi\gamma}\over 6} {k^2+3\gamma^2\over
(k^2+\gamma^2)^2}
\end{eqnarray}
while from the bubble sum in Figure 4b+c we find
\begin{eqnarray}
H_{E1}^{b+c}({}^1P-{}^1S)&=&{es_0\lambda_t\over 2}{1\over 3}\left(
{C_{0s}\over 1-C_{0s}G_0(k)}\right)\int{d^3q\over (2\pi)^3}
G_0(\vec{r}=0,\vec{q})\vec{\nabla}_{\vec{q}}\cdot\left(\vec{q}\psi_d(q)
\right)\nonumber\\
&=&{es_0M\sqrt{8\pi\gamma}\lambda_t\over 6}\left(
{C_{0s}\over 1-C_{0s}G_0(k)}\right)\nonumber\\
&\times&\int{d^3q\over (2\pi)^3} {q^2+3\gamma^2\over
(q^2+\gamma^2)^2({k^2\over M}-{q^2\over M}+i\epsilon)}
\end{eqnarray}
Here the integral may be evaluated via
\begin{eqnarray}
\int{d^3q\over (2\pi)^3} {q^2+3\gamma^2\over
(q^2+\gamma^2)^2({k^2\over M}-{q^2\over M}+i\epsilon)}&=&
(1-2\gamma^2{d\over d\gamma^2}){1\over (k^2+\gamma^2)}
(G_0(k)-G_0(i\gamma))\nonumber\\
&=&{1\over 4\pi}{2\gamma-ik\over (\gamma-ik)^2}
\end{eqnarray}
Summing the two results we find
\begin{eqnarray}
H_{E1}({}^1P-{}^1S)&=&{es_0\sqrt{8\pi\gamma}\lambda_t\over
2}{1\over 3(k^2+\gamma^2)^2}
\left(k^2+3\gamma^2-a_s{(2\gamma-ik)(\gamma+ik)^2\over 1+ika_s}
\right)\nonumber\\
&=&{es_0\sqrt{8\pi\gamma}\lambda_t\over 6}
{k^2+3\gamma^2-2\gamma^3a_s\over (k^2+\gamma^2)^2 (1+ika_s)}
\end{eqnarray}
as found using coordinate space methods.

The matrix element $H_{E1}$ also receives contributions from the
E1 amplitude connecting the deuteron wavefunction with the ${}^3P$
mixture of the final state wavefunction mixed into the ${}^1S_0$.
This admixture can be read off from the Green's function
representation of the scattering amplitude as
\begin{equation}
\delta_{{}^3P}\psi_{{}^1S_0}=-im_s(k)(\vec{\sigma}_p-\vec{\sigma}_n)\cdot\vec{\nabla}{e^{ikr}\over
r}
\end{equation}
and leads to an E1 amplitude
\begin{eqnarray}
H_{E1}({}^3P-{}^3S)&=&{es_0\lambda_s^{np}m_s^*(k)\over 2}{1\over
3} \int d^3r\psi_d(r)\vec{r}\cdot\vec{\nabla}{e^{-ikr}\over r}
\psi_{{}^1S_0}^*(kr)\nonumber\\
&-&{es_0\lambda_s^{np}\sqrt{8\pi\gamma}\over 6}\left({a_s\over
1-ika_s}\right)
\int_0^\infty dr (1+ikr)e^{-(\gamma+ik)r}\nonumber\\
&=&-{es_0\lambda_s^{np}\sqrt{8\pi\gamma}\over 6}\left({a_s\over
1-ika_s}\right)
{\gamma+2ik\over (\gamma+ik)^2}\nonumber\\
&=&-{es_0\lambda_s^{np}a_s\sqrt{8\pi\gamma}e^{-i{\rm
tan}^{-1}ka_s} \over 6(k^2+\gamma^2)^2\sqrt{1+k^2a_s^2}}
(\gamma(\gamma^2+3k^2)-2ik^3)
\end{eqnarray}
Equivalently we can use EFT techniques.  In this case there is no
analog of Figure 4a. For the remaining diagrams, however, we find
\begin{eqnarray}
H_{E1}^{b+c}({}^3P-{}^3S)&=&{es_0\lambda_s^{np}\over 2}{1\over 3}
\left({C_{0s}\over 1-C_{0s}G_0(k)}\right)\nonumber\\
&\times&\int{d^3q\over (2\pi)^3}
\psi_d(q)\vec{\nabla}_{\vec{q}}\cdot\vec{q}G_0({\vec{r}=0},\vec{q})\nonumber\\
&=&{es_0\lambda_s^{np}\sqrt{8\pi\gamma}\over 6}
\left({C_{0s}\over 1-C_{0s}G_0(k)}\right)\nonumber\\
&\times& \left(1+\gamma^2{d\over d\gamma^2}\right){2\over
k^2+\gamma^2}
(G_0^*(k)-G_0^*(-i\gamma))\nonumber\\
&=&{es_0\lambda_s^{np}\over 6} \left({C_{0s}\over
1-C_{0s}G_0(k)}\right){M\over 4\pi}{\gamma(\gamma^2+3k^2)
-2ik^3\over (\gamma^2+k^2)^2}
\end{eqnarray}
{\it i.e.},
\begin{equation}
H_{E1}({}^3P-{}^3S)
=-{es_0\lambda_s^{np}\sqrt{8\pi\gamma}e^{-i{\rm tan}^{-1}ka_s}
\over 6(k^2+\gamma^2)^2\sqrt{1+k^2a_s^2}}
a_s[(\gamma^2+3k^2)\gamma-2ik^3]
\end{equation}
as found in coordinate space.

The full matrix element is then found by combining the singlet and
triplet mixing contributions---
\begin{eqnarray}
H_{E1}&=&H_{E1}({}^3P-{}^3S)+H_{E1}({}^1P-{}^1S)\nonumber\\
&=&{es_0\sqrt{8\pi\gamma}e^{-i\delta_s}\over 6\sqrt{1+k^2a_s^2}
(k^2+\gamma^2)^2}\nonumber\\
&\times&\left[\lambda_t(k^2+3\gamma^2-2a_s\gamma^3)+\lambda_s^{np}\gamma
a_s [(\gamma^2+3k^2)\gamma-2ik^3])\right]
\end{eqnarray}

In the case of the PV E1 matrix element $S_{E1}$ the calculation
appears to be nearly identical, except for the feature that now
the spin triplet final state is involved, so that the calculation
already performed in the case of $H_{E1}$ can be taken over
directly provided that we make the substitutions
$\lambda_s^{np},\lambda_t\rightarrow \rho_t,\,\,a_s\rightarrow
a_t$. The result is found then to be
\begin{equation}
S_{E1}={es_0\sqrt{8\pi\gamma}e^{-i{\rm tan}^{-1}ka_t}\rho_t \over
6\sqrt{1+k^2a_t^2} (k^2+\gamma^2)^2}
\left[(k^2+3\gamma^2-2a_t\gamma^3)+a_t[\gamma
(\gamma^2+3k^2)-2ik^3])\right]\label{eq:seo}
\end{equation}
At the level of approximation we are working we can identify $a_t$
with $1/\gamma$ so that Eq. \ref{eq:seo} becomes
\begin{equation}
S_{E1}={es_0\sqrt{8\pi\gamma}e^{-i{\rm tan}^{-1}{k\over
\gamma}}\rho_t (\gamma^2+2k^2-i{k^3\over \gamma}) \over
3\sqrt{1+{k^2\over \gamma^2}} (k^2+\gamma^2)^2}
\end{equation}

Now consider how to detect these PV amplitudes. The parity
violating electric dipole amplitude $H_{E1}$ can be measured by
looking at the circular polarization which results from
unpolarized radiative capture at threshold or by the asymmetry
resulting from the scattering of polarized photons in
photodisintegration. At threshold, we have for photons of
positive/negative helicity
\begin{equation}
{\rm Amp}_\pm=(\pm
G_{M1V}+H_{E1})\hat{\epsilon}_\gamma\cdot\chi^\dagger_s
(\vec{\sigma}_p-\vec{\sigma}_n)\chi_t+S_{E1}
\hat{\epsilon}_\gamma\cdot\chi^\dagger_t
(\vec{\sigma}_p+\vec{\sigma}_n)\chi_t
\end{equation}
and the corresponding cross sections are found to be
\begin{equation}
\sigma_\pm(np\rightarrow d\gamma)={s_0\over 2\pi|\vec{v}_{rel}|}
|\mp G_{M1V}+H_{E1}|^2+{\cal O}(S_{E1}^2)
\end{equation}
Thus the spin-conserving E1 amplitude $S_{E1}$ does {\it not}
interfere with the leading M1 and the circular polarization is
given by
\begin{eqnarray}
P_\gamma&=&{\sigma_+-\sigma_-\over
\sigma_++\sigma_-}=-{2H_{E1}\over G_{M1V}}=-{4M\over 3\mu_V
(1-\gamma
a_s)(k^2+\gamma^2)}\left[\lambda_t(k^2+3\gamma^2-2a_s\gamma^3)
\right.\nonumber\\
&+&\left.\lambda_s^{np}\gamma a_s(\gamma^2+3k^2))\right]
\end{eqnarray}
A bit of thought makes it clear that this is also the asymmetry
parameter between right- and left-handed circularly polarized
cross sections in the photodisintegration reaction $\gamma_\pm
d\rightarrow np$, and so we have the usual identity between
polarization and asymmetry which is guaranteed by time reversal
invariance
$$P_\gamma(np\rightarrow d\vec{\gamma})=A_\gamma(\vec{\gamma}d\rightarrow np)$$

In order to gain sensitivity to the matrix element $S_{E1}$ one
must use polarized neutrons.  In this case the appropriate trace
is found to be
\begin{equation}
{\rm Tr}{1\over 4}(3+\vec{\sigma}_1\cdot\vec{\sigma}_2) T{1\over
2}(1+\sigma_n\cdot\hat{n})T^\dagger=4|G_{M1V}|^2 +8{\rm
Re}\,G_{M1V}^*S_{E1}\hat{n}\cdot\hat{s}_\gamma
\end{equation}
In this case $H_{E1}$ does not interfere with $G_{M1V}$ and the
corresponding front-back photon asymmetry is
\begin{equation}
A_\gamma={2{\rm Re}G_{M1V}^*S_{E1}\over
|G_{M1V}|^2}=-{8M\rho_t\over 3\mu_V(1-\gamma
a_s)}\left({\gamma^2+2k^2\over \gamma^2+k^2}\right)
\end{equation}
In principle then precise experiments measuring the circular
polarization and photon asymmetry in thermal neutron capture on
protons can produce the remaining two low energy parity violating
parameters $\lambda_t^{np}$ and $\rho_t$ which we seek.  At the
present time only upper limits exist, however. In the case of the
circular polarization we have the number from a Gatchina
measurement\cite{gat}
\begin{equation}
P_\gamma=(1.8\pm 1.8)\times 10^{-7}
\end{equation}
while in the case of the asymmetry we have
\begin{equation}
A_\gamma=(-1.5\pm 4.7)\times 10^{-8}
\end{equation}
from a Grenoble experiment\cite{gre}.  This situation should soon
change, as a new high precision asymmetry measurement at LANL is
being run which seeks to improve the previous precision by an
order of magnitude\cite{lnl}.

\begin{center}
{\bf Appendix}
\end{center}

Of course, as we move above threshold we must also include the
parity-violating M1 matrix elements, which interfere with the
leading E1 amplitude and are of two types. The first is the M1
amplitude which connects the ${}^1P_1$ admixture of the deuteron
with the ${}^3P$ np scattering state as well as the M1 amplitude
connecting the ${}^1S_0$ admixture of the final ${}^3P$ scattering
state with the deuteron ground state. For the former we
have\footnote{For simplicity here we include only the dominant
isovector M1 amplitude.  A complete discussion should also include
the corresponding isoscalar M1 transition.}
\begin{equation}
{\rm
Amp}=J_{M1}^a\chi_t^\dagger(\vec{\sigma}_1-\vec{\sigma}_2)\cdot
\hat{\epsilon}_\gamma\times\hat{s}_\gamma(\vec{\sigma}_1-\vec{\sigma}_2)\cdot
\vec{k}\chi_t
\end{equation}
where
\begin{equation}
J_{M1}^a={es_0\mu_V\over 4M} {\lambda_t\over k^2}\int
d^3re^{-i\vec{k}\cdot\vec{r}}
\vec{k}\cdot\vec{\nabla}\psi_d(r)={es_0\mu_V\over 4M}
{\lambda_t\sqrt{8\pi\gamma}\over (k^2+\gamma^2)}
\end{equation}
while for the latter we find
\begin{equation}
{\rm
Amp}=J_{M1}^b\chi_t^\dagger(\vec{\sigma}_1-\vec{\sigma}_2)\cdot
\vec{k}(\vec{\sigma}_1-\vec{\sigma}_2)\cdot
\hat{\epsilon}_\gamma\times\hat{s}_\gamma\chi_t
\end{equation}
where
\begin{equation}
J_{M1}^b={es_0\mu_V\over 4M} \lambda_s^{np}m_s^*(k)\int
d^3r{1\over r}e^{-ikr}\psi_d(r)=-{es_0\mu_V\over 4M}
{\lambda_s^{np}a_s\sqrt{8\pi\gamma}\over (1-ika_s)(\gamma+ik)}
\end{equation}

A second category of PV M1 amplitudes involves that which connects
the ${}^3P_1$ piece of the deuteron wavefunction with the
${}^1P_1$ or ${}^3P$ np scattering states as well as the M1
amplitude connecting the ${}^3S_1$ or ${}^1S_0$ admixture of the
final ${}^3P$ scattering state with the deuteron ground state. For
the former we have
\begin{equation}
K_{M1}^a\chi^\dagger_s[(\vec{\sigma}_1-\vec{\sigma}_2)\cdot\hat{\epsilon}_\gamma
\times\hat{s}_\gamma(\vec{\sigma}_1+\vec{\sigma}_2)\cdot\vec{k}\chi_t
\end{equation}
\begin{equation}
L_{M1}^a\chi^\dagger_t[(\vec{\sigma}_1+\vec{\sigma}_2)\cdot\hat{\epsilon}_\gamma
\times\hat{s}_\gamma(\vec{\sigma}_1+\vec{\sigma}_2)\cdot\vec{k}\chi_t
\end{equation}
where
\begin{equation}
K_{M1}^a={es_0\mu_V\over 4M}{\rho_t\over k^2}\int
d^3re^{-i\vec{k}\cdot\vec{r}}\vec{k}\cdot\vec{\nabla}\psi_d(r)={es_0\mu_V\over
4M}{\rho_t\sqrt{8\pi\gamma}\over k^2+\gamma^2}
\end{equation}
\begin{equation}
L_{M1}^a={es_0\mu_S\over 4M}{\rho_t\over k^2}\int
d^3re^{-i\vec{k}\cdot\vec{r}}\vec{k}\cdot\vec{\nabla}\psi_d(r)={es_0\mu_S\over
4M}{\rho_t\sqrt{8\pi\gamma}\over k^2+\gamma^2}
\end{equation}
while for the latter we have
\begin{equation}
{\rm
Amp}=K_{M1}^b\chi_s^\dagger(\vec{\sigma}_1+\vec{\sigma}_2)\cdot
\vec{k}(\vec{\sigma}_1-\vec{\sigma}_2)\cdot
\hat{\epsilon}_\gamma\times\hat{s}_\gamma\chi_t
\end{equation}
\begin{equation}
{\rm
Amp}=L_{M1}^b\chi_t^\dagger(\vec{\sigma}_1+\vec{\sigma}_2)\cdot
\vec{k}(\vec{\sigma}_1+\vec{\sigma}_2)\cdot
\hat{\epsilon}_\gamma\times\hat{s}_\gamma\chi_t
\end{equation}
where
\begin{equation}
K_{M1}^b={es_0\mu_V\over 4M}\rho_tm_t^*(k)\int d^3r{1\over
r}e^{-ikr}\psi_d(r)=-{es_0\mu_V\over 4M}
{\rho_ta_t\sqrt{8\pi\gamma}\over (1-ika_t)(\gamma+ik)}
\end{equation}
\begin{equation}
L_{M1}^b={es_0\mu_S\over 4M}\rho_tm_t^*(k)\int d^3r{1\over
r}e^{-ikr}\psi_d(r)=-{es_0\mu_S\over 4M}
{\rho_ta_t\sqrt{8\pi\gamma}\over (1-ika_t)(\gamma+ik)}
\end{equation}
To this order we can use $a_t\simeq 1/\gamma$, so that
\begin{equation}
K_{M1}^b=-{es_0\mu_V\over 4M}{\rho_t\sqrt{8\pi\gamma}\over
k^2+\gamma^2}
\end{equation}
\begin{equation}
L_{M1}^b=-{es_0\mu_S\over 4M}{\rho_t\sqrt{8\pi\gamma}\over
k^2+\gamma^2}
\end{equation}
We see then that this piece of the M1 amplitude has the form
\begin{equation}
{\rm
Amp}=2i(\hat{\epsilon}_\gamma\times\hat{s}_\gamma)\times\vec{k}\cdot\chi_f^\dagger[
K_{M1}^a(\vec{\sigma}_1-\vec{\sigma}_2)+L_{M1}^a(\vec{\sigma}_1+\vec{\sigma}_2)]\chi_t
\end{equation}

The relevant traces here are
\begin{eqnarray} &&{\rm
Tr}{1\over 4}(3+\vec{\sigma}_1\cdot\vec{\sigma}_2)\left(
J_{M1}^a(\vec{\sigma}_1-\vec{\sigma}_2)\cdot
\hat{\epsilon}_\gamma\times\hat{s}_\gamma(\vec{\sigma}_1-\vec{\sigma}_2)\cdot
\vec{k}\right.\nonumber\\
&+&\left.J_{M1}^b(\vec{\sigma}_1-\vec{\sigma}_2)\cdot
\vec{k}(\vec{\sigma}_1-\vec{\sigma}_2)\cdot
\hat{\epsilon}_\gamma\times\hat{s}_\gamma\right){1\over 2
}(1+\vec{\sigma}_2\cdot\hat{n})\nonumber\\
&=&2(J_{M1}^a+J_{M1}^b)
\hat{\epsilon}_\gamma\times\hat{s}_\gamma\cdot\vec{k}+2i(J_{M1}^a-J_{M1}^b)\hat{n}\cdot(\hat{\epsilon}_\gamma
\times\hat{s}_\gamma)\times\vec{k}
\end{eqnarray}
\begin{eqnarray}
&&{\rm Tr}{1\over 4}(3+\vec{\sigma}_1\cdot\vec{\sigma}_2)\left(
K_{M1}^a(\vec{\sigma}_1-\vec{\sigma}_2)\cdot
\hat{\epsilon}_\gamma\times\hat{s}_\gamma(\vec{\sigma}_1+\vec{\sigma}_2)\cdot
\vec{k}\right.\nonumber\\
&+&\left.K_{M1}^b(\vec{\sigma}_1+\vec{\sigma}_2)\cdot
\vec{k}(\vec{\sigma}_1-\vec{\sigma}_2)\cdot
\hat{\epsilon}_\gamma\times\hat{s}_\gamma\right){1\over 2
}(1+\vec{\sigma}_2\cdot\hat{n})\nonumber\\
&=&2(L_{M1}^a+L_{M1}^b)
\hat{\epsilon}_\gamma\times\hat{s}_\gamma\cdot\vec{k}
+2i(L_{M1}^a-L_{M1}^b)\hat{n}\cdot(\hat{\epsilon}_\gamma
\times\hat{s}_\gamma)\times\vec{k}\nonumber\\
&=&4iL_{M1}^a\hat{n}\cdot(\hat{\epsilon}_\gamma
\times\hat{s}_\gamma)\times\vec{k}
\end{eqnarray}
\begin{eqnarray}
&&{\rm Tr}{1\over 4}(3+\vec{\sigma}_1\cdot\vec{\sigma}_2)\left(
L_{M1}^a(\vec{\sigma}_1+\vec{\sigma}_2)\cdot
\hat{\epsilon}_\gamma\times\hat{s}_\gamma(\vec{\sigma}_1+\vec{\sigma}_2)\cdot
\vec{k}\right.\nonumber\\
&+&\left.L_{M1}^b(\vec{\sigma}_1+\vec{\sigma}_2)\cdot
\vec{k}(\vec{\sigma}_1+\vec{\sigma}_2)\cdot
\hat{\epsilon}_\gamma\times\hat{s}_\gamma\right){1\over 2
}(1+\vec{\sigma}_2\cdot\hat{n})\nonumber\\
&=&2(L_{M1}^a+L_{M1}^b)
\hat{\epsilon}_\gamma\times\hat{s}_\gamma\cdot\vec{k}
+2i(L_{M1}^a-L_{M1}^b)\hat{n}\cdot(\hat{\epsilon}_\gamma
\times\hat{s}_\gamma)\times\vec{k}\nonumber\\
&=&4iL_{M1}^a\hat{n}\cdot(\hat{\epsilon}_\gamma
\times\hat{s}_\gamma)\times\vec{k}
\end{eqnarray}
and the corresponding contribution to the cross section for
photodisintegration by photons of differing helicity is
\begin{eqnarray}
\sigma_\pm&=&{M^2k^3\over
12\pi(\gamma^2+k^2)}\left(|G_{E1}|^2\pm{8\over 3}{\rm
Re}G_{E1}^*(J_{M1}^a+j_{M1}^b)\right)\nonumber\\
&=&{8\pi\gamma k^3\alpha\over 3(k^2+\gamma^2)^3}\pm{16\pi\gamma
k^3\alpha\mu_V\over
9(k^2+\gamma^2)}\left(\lambda_t-\lambda_s^{np}a_s{\gamma+k^2a_s\over
1+k^2a_s^2}\right)
\end{eqnarray}

Note that there is only sensitivity to the couplings $J_{M1}$
here. In order to have sensitivity to the couplings
$K_{M1},\,L_{M1}$ we must look at the E1 contribution to the cross
section for the radiative capture of polarized neutrons--
\begin{eqnarray}
{d\sigma\over d\Omega_{\hat{s}_\gamma}}&=&{\gamma^2+k^2\over
32\pi^2|\vec{v}_{rel}|M}\left(3|G_{E1}|^2(k^2-(\vec{k}\cdot\hat{s}_\gamma)^2)
+8{\rm Re}G_{E1}^*(K_{M1}^a+L_{M1}^a)\hat{s}_\gamma\cdot\vec{k}
\hat{s}_\gamma\times\vec{k}\cdot\hat{n}\right)\nonumber\\
&=&{3\alpha\gamma\over
M^3|\vec{v}_{rel}|}{k^2-(\vec{k}\cdot\hat{s}_\gamma)^2\over
k^2+\gamma^2}+{2\alpha\gamma(\mu_V+\mu_S)\rho_t\over
M^4|\vec{v}_{rel}|}\hat{s}_\gamma\cdot\vec{k}
\hat{s}_\gamma\times\vec{k}\cdot\hat{n}
\end{eqnarray}

Careful analysis of above threshold experiments should include
these NLO corrections to the analysis.

\section{Connecting with Theory} Having a form of the weak
parity-violating potential $V^{PNC}(r)$ it is, of course,
essential to complete the process by connecting with the
S-matrix---{\it i.e.}, expressing the phenomenological parameters
$\lambda_i,\,\rho_t$ defined in Eq. \ref{eq:eff} in terms of the
fundamental ones---$C_i,\,\tilde{C}_i$ defined in Eq.
\ref{eq:sht}.  This is a major undertaking and should involve the
latest and best NN wavefunctions such as Argonne V18.  The work is
underway, but it will be some time until this process is
completed.  Even after this connection has been completed, the
results will be numerical in form.  However, it is very useful to
have an analytic form by which to understand the basic physics of
this transformation and by which to make simple numerical
estimates.  For this purpose we shall employ simple
phenomenological NN wavefunctions, as described below.

Examination of the scattering matrix Eq. \ref{eq:tpx} reveals that
the parameters $\lambda_{s,t}$ are associated with the
(short-distance) component while $\rho_t$ contains contributions
from the both (long-distance) pion exchange as well as short
distance effects. In the former case, since the interaction is
short ranged we can use this feature in order to simplify the
analysis. Thus, we can determine the shift in the deuteron
wavefunction associated with parity violation by demanding
orthogonality with the $^3S_1$ scattering state, which yields,
using the simple asymptotic form of the bound state
wavefunction\cite{khk},\cite{khr}
\begin{equation}
\psi_d(r)=\left[1+\rho_t(\vec{\sigma}_p+\vec{\sigma}_n)\cdot
-i\vec{\nabla} +\lambda_t(\vec{\sigma}_p-\vec{\sigma}_n)\cdot
-i\vec{\nabla})\right] \sqrt{\gamma\over 2\pi}{1\over r}e^{-\gamma
r}\label{eq:deu3}
\end{equation}
where $\gamma^2/M=2.23$ MeV is the deuteron binding energy. Now
the shift generated by $V^{PV}(r)$ is found to
be\cite{khk},\cite{khr}
\begin{eqnarray}
\delta\psi_d(\vec{r}) &\simeq&\int
d^3r'G(\vec{r},\vec{r}')V^{PV}(\vec{r}')
\psi_d(r')\nonumber\\
&=&-{M\over 4\pi}\int d^3r'{e^{-\gamma|\vec{r}-\vec{r}'|}
\over |\vec{r}-\vec{r}'|}V^{PV}(\vec{r}')\psi_d(r')\nonumber\\
&\simeq& {M\over 4\pi}\vec{\nabla}\left({e^{-\gamma r}\over
r}\right)\cdot \int
d^3r'\vec{r}'V^{PV}(\vec{r}')\psi_d(r')\label{eq:shr}
\end{eqnarray}
where the last step is permitted by the short range of
$V^{PV}(\vec{r}')$.  Comparing Eqs. \ref{eq:shr} and \ref{eq:deu3}
yields then the identification
\begin{equation}
\sqrt{\gamma\over 2\pi}\lambda_t\chi_t\equiv i{M\over
16\pi}\xi_0^\dagger\int d^3r'
(\vec{\sigma}_1-\vec{\sigma}_2)\cdot\vec{r}'V^{PV}
(\vec{r}')\psi_d(r')\chi_t\xi_0 \label{eq:elt}
\end{equation}
where we have included the normalized isospin wavefunction $\xi_0$
since the potential involves $\vec{\tau}_1,\vec{\tau}_2.$  When
operating on such an isosinglet np state the PV potential can be
written as
\begin{eqnarray}
V^{PV}(\vec{r}')&=&{2\over \Lambda_\chi^3}\left[
(C_1-3C_3)(\vec{\sigma}_1-\vec{\sigma}_2)\cdot(-i\vec{\nabla}f_m(r)+2f_m(r)
\cdot-i\vec{\nabla})\right.\nonumber\\
&+&\left.(\tilde{C}_1-3\tilde{C}_3)(\vec{\sigma}_1\times\vec{\sigma}_2)
\cdot\vec{\nabla}f_m(r)\right]
\end{eqnarray}
where $f_m(r)$ is the Yukawa form
$$f_m(r)={m^2e^{-mr}\over 4\pi r}$$
defined in Eq. \ref{eq:yuk}.  Using the identity
\begin{equation}
(\vec{\sigma}_1\times\vec{\sigma}_2){1\over 2}
(1+\vec{\sigma}_1\cdot\vec{\sigma}_2)=i(\vec{\sigma}_1
-\vec{\sigma}_2 )
\end{equation}
Eq. \ref{eq:elt} becomes
\begin{eqnarray}
\sqrt{\gamma\over 2\pi}\lambda_t\chi_t&\simeq&{2M\over
16\pi\Lambda_\chi^3}{4\pi\over
3}(\vec{\sigma}_1-\vec{\sigma}_2)^2\chi_t\int_0^\infty
drr^3\nonumber\\
&\times&\left[-2(3C_3-C_1)f_m(r){d\psi_d(r)\over
dr}+(3\tilde{C}_3-3C_3-\tilde{C}_1+C_1)
{df_m(r)\over dr}\psi_d(r)\right]\nonumber\\
&=&\sqrt{\gamma\over 2\pi}\cdot 4\chi_t{1\over 12}{2Mm^2\over
4\pi\Lambda_\chi^3}\left(
{2m(6C_3-3\tilde{C}_3-2C_1+\tilde{C}_1)+\gamma(15C_3-3\tilde{C}_3-5C_1+\tilde{C}_1)\over
(\gamma+m)^2}\right)\nonumber\\
\quad
\end{eqnarray}
or
\begin{equation}
\lambda_t\simeq {Mm^2\over 6\pi\Lambda_\chi^3}\left(
{2m(6C_3-3\tilde{C}_3-2C_1+\tilde{C}_1)+\gamma(15C_3-3\tilde{C}_3-5C_1+\tilde{C}_1)\over
(\gamma+m)^2}\right)
\end{equation}

In order to determine the singlet parameter $\lambda_s^{np}$, we
must use the ${}^1S_0$ np-scattering wavefunction instead of the
deuteron, but the procedure is similar,
yielding\cite{khk},\cite{khr}
\begin{equation}
d_s^{np}(k)\chi_s\equiv{M\over 48\pi}\xi_1^\dagger\int d^3r'
(\vec{\sigma}_1-\vec{\sigma}_2)\cdot\vec{r}'V^{PV}(\vec{r}')
\psi_{{}^1S_0}(r')\chi_s\xi_1 \label{eq:els}
\end{equation}
and we can proceed similarly.  In this case the potential becomes
\begin{eqnarray}
V^{PV}(\vec{r}')&=&{2\over \Lambda_\chi^3}\left[(C_1+C_3+{1\over
6}C_5)
(\vec{\sigma}_1-\vec{\sigma}_2)\cdot(-i\vec{\nabla}f_m(r)+2f_m(r)
\cdot-i\vec{\nabla})\right.\nonumber\\
&+&\left.(\tilde{C}_1+\tilde{C}_3+{1\over
6}\tilde{C}_5)(\vec{\sigma}_1\times\vec{\sigma}_2)\cdot
\vec{\nabla}f_m(r)\right]
\end{eqnarray}
and Eq. \ref{eq:els} is found to have the form
\begin{eqnarray}
d_s^{np}(k)\chi_s&=&{2M\over 48\pi\Lambda_\chi^3}{4\pi\over 3}
(\vec{\sigma}_1-\vec{\sigma}_2)^2\chi_s\int_0^\infty dr{r}^3\left[
\right.\nonumber\\
&\times&\left.2(C_1+C_3+4C_5)f_m(r){d\psi_{{}^1S_0}(r)\over
dr}\right.\nonumber\\
&+&\left.(C_1+\tilde{C}_1+C_3+\tilde{C}_3+4(C_5+\tilde{C}_5))
{df_m(r)\over dr}\psi_{{}^1S_0}(r)\right]\nonumber\\
 &=&-12\chi_s{1\over 36}{2Mm^2\over
4\pi\Lambda_\chi^3}e^{i\delta_s}\left[{1\over
(k^2+m^2)^2}\right.\nonumber\\
&\times&\left.\left(\cos\delta_s(4k^2(C_1+C_3+4C_5)+(C_1+\tilde{C}_1+C_3+\tilde{C}_3
+4(C_5+\tilde{C}_5))(k^2+3m^2))
\right.\right.\nonumber\\
&+&\left.\left.{2m\over
k}\sin\delta_s((C_1+C_3+4C_5)(m^2+3k^2)+(C_1+\tilde{C}_1+C_3+\tilde{C}_3+
4(C_5+\tilde{C}_5))m^2)\right) \right)\nonumber\\
\quad
\end{eqnarray}
which, in the limit as $k\rightarrow 0$, yields the predicted
value for $\lambda_s^{np}$---
\begin{eqnarray}
\lambda_s^{np}&=&-{1\over a_s^{np}}\lim_{k\rightarrow
0}d_s^{np}(k)= {M\over 6\pi a_s^{np}\Lambda_\chi^3}\left[
3(C_1+\tilde{C}_1+C_3+\tilde{C}_3+4(C_5+\tilde{C}_5))\right.\nonumber\\
&-&\left.2ma_s^{np}(2C_1+\tilde{C}_1+2C_3+\tilde{C}_3+4(2C_5+\tilde{C}_5)))\right]
\end{eqnarray}
Similarly, we may identify
\begin{eqnarray}
\lambda_s^{pp}&=&-{1\over a_s^{pp}}\lim_{k\rightarrow
0}d_s^{pp}(k)= {M\over 6\pi a_s^{pp}\Lambda_\chi^3}\left[
3(C_1+\tilde{C}_1+C_2+\tilde{C}_2+C_3+\tilde{C}_3+C_4+\tilde{C}_4-2(C_5+\tilde{C}_5))\right.\nonumber\\
&-&\left.2ma_s^{pp}(2C_1+\tilde{C}_1+2C_2+\tilde{C}_2+2C_3+\tilde{C}_3+2C_4+\tilde{C}_4-
2(2C_5+\tilde{C}_5)))\right]\nonumber\\
\lambda_s^{nn}&=&-{1\over a_s^{nn}}\lim_{k\rightarrow
0}d_s^{nn}(k)= {M\over 6\pi a_s^{nn}\Lambda_\chi^3}\left[
3(C_1+\tilde{C}_1-C_2-\tilde{C}_2+C_3+\tilde{C}_3-C_4-\tilde{C}_4-2(C_5+\tilde{C}_5))\right.\nonumber\\
&-&\left.2ma_s^{nn}(2C_1+\tilde{C}_1-2C_2-\tilde{C}_2+2C_3+\tilde{C}_3-2C_4-\tilde{C}_4-
2(2C_5+\tilde{C}_5)))\right]\nonumber\\
\quad
\end{eqnarray}
In order to evaluate the spin-conserving amplitude $\rho_t$, we
shall assume dominance of the long range pion component. The shift
in the deuteron wavefunction is given by
\begin{eqnarray}
\delta \psi_d(\vec{r})&=&\xi_0^\dagger\int
d^3r'G_0(\vec{r},\vec{r}')V^{PV}_{\pi}(\vec{r}')
\psi_d(r')\nonumber\\
&=&-{M\over 4\pi}\xi_0^\dagger\int
d^3r'{e^{-\gamma|\vec{r}-\vec{r}'|} \over
|\vec{r}-\vec{r}'|}V^{PV}_{\pi}(\vec{r}')\psi_d(r')\chi_t\xi_0
\end{eqnarray}
but now with\footnote{Here we have used the identity
\begin{equation}
(\vec{\tau}_1\times\vec{\tau}_2)=-i(\vec{\tau_1}-\vec{\tau}_2){1\over
2} (1+\vec{\tau}_1\cdot\vec{\tau}_2)
\end{equation}}
\begin{equation}
V^{PV}_\pi(\vec{r})={h_\pi g_{\pi NN}\over
\sqrt{2}Mm_\pi^2}{1\over 2}(\tau_1-\tau_2)_z
(\vec{\sigma}_1+\vec{\sigma}_2)\cdot-i\vec{\nabla}f_\pi(r)
\end{equation}
Of course, the meson which is exchanged is the pion so the short
range assumption which permitted the replacement in Eq.
\ref{eq:shr} is not valid and we must perform the integration
exactly.  This process is straightforward but tedious\cite{des}.
Nevertheless, we can get a rough estimate by making a ``heavy
pion'' approximation, whereby we can identify the constant
$\rho_t$ via
\begin{equation}
\sqrt{\gamma\over 2\pi}\rho_t\chi_t\approx -i{M\over 32\pi}\int
d^3r' (\vec{\sigma}_1+\vec{\sigma}_2)\cdot\vec{r}'V^{PV}_{\pi}
(\vec{r}')\psi_d(r')\chi_t\xi_0
\end{equation}
which leads to\cite{dpl}
\begin{eqnarray}
\sqrt{\gamma\over 2\pi}\rho_t\chi_t&\approx&{1\over
32\pi}{4\pi\over 3} (\vec{\sigma}_1+\vec{\sigma}_2)^2\chi_t3{h_\pi
g_{\pi NN}
\over \sqrt{2}}\int_0^\infty drr^3{df_\pi(r)\over dr}\psi_d(r)\nonumber\\
&=&\sqrt{\gamma\over 2\pi}8\chi_t{1\over 96\pi}{h_\pi g_{\pi
NN}\over \sqrt{2}} {\gamma+2m_\pi\over (\gamma+m_\pi)^2}
\end{eqnarray}
We find then the prediction
\begin{equation}
\rho_t={g_{\pi NN}\over 12\sqrt{2}\pi}{\gamma+2m_\pi\over
(\gamma+m_\pi)^2}h_\pi
\end{equation}

At this point it is useful to obtain rough numerical estimates.
This can be done by use of the numerical estimates given in Table
2.  To make things tractable, we shall use the best values given
therein.  Since we are after only rough estimates and since the
best values assume the DDH relationship---Eq. \ref{eq:ddhr}
between the tilde- and non-tilde- quantities, we shall express our
results in terms of only the non-tilde numbers.  Of course, a
future complete evaluation should include the full dependence.  Of
course, these predictions are only within a model, but they has
the advantage of allowing connection with previous theoretical
estimates.  In this way, we find the predictions
\begin{eqnarray}
\lambda_t&=&\left[-0.092C_3-0.014C_1\right]m_\pi^{-1}\nonumber\\
\lambda_s^{np}&=&\left[-0.087(C_3+4C_5)-0.037C_1\right]
m_\pi^{-1}\nonumber\\
\lambda_s^{pp}&=&\left[-0.087(C_3+C_4-2C_5)
-0.037(C_1+C_2)\right]m_\pi^{-1}\nonumber\\
\lambda_s^{nn}&=&\left[-0.087(C_3-C_4-2C_5)
-0.037(C_1-C_2)\right]m_\pi^{-1}\nonumber\\
\rho_t&=& 0.346 h_\pi m_\pi^{-1}
\end{eqnarray}
so that, using the relations Eq. \ref{eq:rel} and the best values
from Table 1 we estimate
\begin{eqnarray}
\lambda_t&=&-2.39\times 10^{-7}m_\pi^{-1}=-3.41\times
10^{-7}\,\,{\rm fm}\nonumber\\
\lambda_s^{np}&=&-1.12\times 10^{-7}m_\pi^{-1}=-1.60\times
10^{-7}\,\,
{\rm fm}\nonumber\\
\lambda_s^{pp}&=&-3.58\times 10^{-7}\,\,m_\pi^{-1}=-5.22\times
10^{-7}\,\,{\rm fm}\\
\lambda_s^{nn}&=&-2.97\times 10^{-7}\,\,m_\pi^{-1}=-4.33\times
10^{-7}\,\,{\rm fm}\nonumber\\
\rho_t&=&1.50\times 10^{-7}\,\,m_\pi^{-1}=2.14\times
10^{-7}\,\,{\rm fm}
\end{eqnarray}

At this point we note, however, that $\lambda_s^{pp}$ is an order
of magnitude larger than the experimentally determined number, Eq.
\ref{eq:ppe}.  The problem here is not with the couplings but with
an important piece of physics which has thus far been
neglected---short distance effects. There are two issues here. One
is that the deuteron and NN wavefunctions should be modified at
short distances from the simple asymptotic form used up until this
point in order to account for finite size effects. The second is
the well-known feature of the Jastrow correlations that suppress
the nucleon-nucleon wavefunction at short distance.

In order to deal approximately with the short distance properties
of the deuteron wavefunction, we modify the exponential form to
become constant inside the deuteron radius
$R$\cite{khk},\cite{khr}
\begin{equation}
\sqrt{\gamma\over 2\pi}{1\over r}e^{-\gamma r} \rightarrow
N\left\{\begin{array}{cc}
{1\over R}e^{-\gamma R}& r\leq R\\
{1\over r}e^{-\gamma r}& r>R
\end{array}\right.
\end{equation}
where
$$N=\sqrt{\gamma\over 2\pi}{\exp\gamma R\over \sqrt{1+{2\over 3}\gamma R}}$$
is the modified normalization factor and we use R=1.6 fm.  For the
NN wavefunction we use
\begin{equation}
\psi_{{}^1S_0}(r)=\left\{\begin{array}{cc}
A{\sin\sqrt{p^2+p_0^2}r\over \sqrt{p^2+p_0^2}r}& r\leq r_s\\
{\sin pr\over pr}-{1\over {1\over a_s}+ip}{e^{ipr}\over r}&r>r_s
\end{array}\right.
\end{equation}
where we choose $r_s=2.73$ fm and $p_or_s=1.5$.  The normalization
constant $A(p)$ is found by requiring continuity of the
wavefunction and its first derivative at $r=r_s$
\begin{equation}
A(p)={\sqrt{p^2+p_0^2}r_s\over \sin\sqrt{p^2+p_0^2}r_s}{\sin pr_s-
pa_s\cos pr_s\over pr_s(1+ipa_s)}
\end{equation}
As to the Jastrow correlations we multiply the wavefunction by the
simple phenomenological form\cite{rib}
\begin{equation}
\phi(r)=1-ce^{-dr^2},\quad{\rm with}\quad c=0.6,\quad d=3\,\,{\rm
fm}^{-2}
\end{equation}
With these modifications we find the much more reasonable values
for the constants $\lambda_s^{{pp},{np}}$ and $\lambda_t$
\begin{eqnarray}
\lambda_s^{pp}&=&\left[-0.011(C_3+C_4-2C_5)-0.004
(C_1+C_2)\right]m_\pi^{-1}\nonumber\\
\lambda_s^{nn}&=&\left[-0.011(C_3-C_4+2C_5)-0.004
(C_1-C_2)\right]m_\pi^{-1}\nonumber\\
\lambda_s^{np}&=&\left[-0.011(C_3+4C_5)-0.004C_1\right]m_\pi^{-1}\nonumber\\
\lambda_t&=&\left[-0.019C_3-0.0003C_1\right]m_\pi^{-1}
\label{eq:cal}
\end{eqnarray}
Using the best values from Table 2 we find then the benchmark
values
\begin{eqnarray}
\lambda_s^{pp}&=&-4.2\times 10^{-8}m_\pi^{-1}= -6.1\times
10^{-8}\,\,{\rm fm}
\nonumber\\
\lambda_s^{nn}&=&-3.6\times 10^{-8}m_\pi^{-1}= -5.3\times
10^{-8}\,\,{\rm fm}
\nonumber\\
\lambda_s^{np}&=&-1.3\times 10^{-8}m_\pi^{-1}= -1.9\times
10^{-8}\,\,{\rm fm}
\nonumber\\
\lambda_t&=&-4.7\times 10^{-8}m_\pi^{-1}=-6.7\times
10^{-8}\,\,{\rm fm}
\end{eqnarray}
Since $\rho_t$ is a long distance effect, we use the same value as
calculated previously as our benchmark number
\begin{equation}
\rho_t=1.50\times 10^{-7}\,\,m_\pi^{-1}=2.14\times 10^{-7}\,\,{\rm
fm}
\end{equation}

Obviously the value of $\lambda_s^{pp}$ is now in much better
agreement with the experimental value Eq. \ref{eq:ppe}.  Of
course, our rough estimate is no substitute for a reliable state
of the art wavefunction evaluation.  This has been done recently
by Carlson et al. and yields, using the Argonne V18
wavefunctions\cite{car}
\begin{equation}
\lambda_s^{pp}=\left[-0.008(C_3+C_4-2C_5)
-0.003(C_1+C_2)\right]m_\pi^{-1}
\end{equation}
in reasonable agreement with the value calculated in Eq.
\ref{eq:cal}. Similar efforts should be directed toward evaluation
of the remaining parameters using the best modern wavefunctions.

We end our brief discussion here, but clearly this was merely a
simplistic model calculation. It is important to complete this
process by using the best contemporary nucleon-nucleon
wavefunctions with the most general EFT potential developed above,
in order to allow the best possible restrictions to be placed on
the unknown counterterms.

\section{Summary}

For nearly fifty years both theorists and experimentalists have
been struggling to obtain an understanding of the parity-violating
nucleon-nucleon interaction and its manifestations in hadronic
parity violation.  Despite a great deal of effort on both fronts,
at the present time there still exists a great deal of confusion
both as to whether the DDH picture is able to explain the data
which exists and even if this is the case as to the size of the
basic weak couplings.  For this reason it has recently been
advocated to employ and effective field theory approach to the low
energy data, which must be describable in terms of five elementary
S-=matrix elements.  Above we have discussed both connection
between these S-matrix elements and observables in the NN system
via simple analytic methods based both on a conventional
wavefunction approach as well as on effective field theory
methods.  While the results are only approximate, they are in
reasonable agreement with those obtained via precision state of
the art nonrelativistic potential calculations and serve we hope
to aid in the understanding of the basic physics of the
parity-violating NN sector.  In this way it is hoped that the
round of experiments which is currently underway can be used to
produce a reliable set of weak couplings, which can in turn be
used both in order to connect with more fundamental theory such as
QCD as well as to provide a solid basis for calculations wherein
such hadronic parity violation is acting.

\begin{center}
{\bf Acknowledgement}
\end{center}

This work was supported in part by the National Science Foundation
under award PHY/02-44801.  I also wish to express my appreciation
to Prof. I.B. Khriplovich, from whose work I extracted many of
these methods.


\begin{thebibliography}{99}
\bibitem{part} Note that we are using here the standard convention in nuclear
physics wherein the scattering length associated with a weak
attractive potential is negative.  It is common in particle
physics to define the scattering length as the negative of that
given here so that a weak attractive potential is associated with
a positive scattering length.
\end{thebibliography}
\end{document}